\documentclass{aa}
\usepackage{aalongtable}
\usepackage{graphicx}
\usepackage{color}
\usepackage{txfonts}
\usepackage{enumerate}
\usepackage{natbib}

\begin{document}

\title{Rotational velocities of A-type stars 
\thanks{Full Table\,\ref{fund_par} is only available in electronic form at CDS via anonymous ftp to \texttt{cdsarc.u-strasbg.fr} (\texttt{130.79.128.5}) or via \texttt{http://cdsweb.u-strasbg.fr/cgi-bin/qcat?J/A+A/}}
\thanks{Appendix is only available in electronic form at \texttt{http://www.aanda.org}}
}
\subtitle{IV. Evolution of rotational velocities}

\author{
J. Zorec\inst{1}
\and 
F. Royer\inst{2}
}

\offprints{Fr\'ed\'eric Royer,\\ \email{frederic.royer@obspm.fr}}
 
\institute{
Institut d'Astrophysique de Paris, UMR 7095 CNRS -- Universit\'e Pierre \&
Marie Curie, 98 bis boulevard Arago, 75014 Paris, France
\and 
GEPI/CNRS UMR 8111, Observatoire de Paris -- Universit\'e Paris Denis Diderot, 5 place Jules Janssen, 92195 Meudon cedex, France
}

\date{Received / Accepted}

\titlerunning{Rotational velocities of A-type stars. IV.}
\authorrunning{Zorec \& Royer}

\abstract
{In previous works of this series, we have shown that late B- and early A-type stars have genuine bimodal distributions of rotational velocities and that late A-type stars lack slow rotators. The distributions of the surface angular velocity ratio $\Omega/\Omega_{\rm crit}$ ($\Omega_{\rm crit}$ is the critical angular velocity) have peculiar shapes according to spectral type groups, which can be caused by evolutionary properties.}
{We aim to review the properties of these rotational velocity distributions in some detail as a function of stellar mass and age.}
{We have gathered \ensuremath{v\sin i}\ for a sample of 2014 B6- to F2-type stars. We have determined the masses and ages for these objects with stellar evolution models. The ($\ensuremath{T_\mathrm{eff}}, \log L/L_{\odot}$)-parameters were determined from the $uvby$--$\beta$ photometry and the HIPPARCOS parallaxes.}
{The velocity distributions show two regimes that depend on the stellar mass. Stars less massive than 2.5\,\ensuremath{M_\odot}\ have a unimodal equatorial velocity distribution and show a monotonical acceleration with age on the main sequence (MS). Stars more massive have a bimodal equatorial velocity distribution. Contrarily to theoretical predictions, the equatorial velocities of stars from about $1.7\,\ensuremath{M_\odot}$ to $3.2\,\ensuremath{M_\odot}$ undergo a strong acceleration in the first third of the MS evolutionary phase, while in the last third of the MS they evolve roughly as if there were no angular momentum redistribution in the external stellar layers. The studied stars might start in the ZAMS not necessarily as rigid rotators, but with a total angular momentum lower than the critical one of rigid rotators. The stars seem to evolve as differential rotators all the way of their MS life span and the variation of the observed rotational velocities proceeds with characteristic time scales $\delta t\approx 0.2\,t_{\rm MS}$, where $t_{\rm MS}$ is the time spent by a star in the MS.}
{}



\maketitle

\section{Introduction}\label{introd}

 A great number of phenomena are involved in establishing the observed surface rotational velocities of stars. There is a first redistribution of angular momentum among the fragments determined by the fragmentation process of protostellar clouds \citep{Bor81,Mao_97}, which need to be dissipated in different ways before they reach the birth-line. This may happen mostly through magnetic braking \citep{MosMon85a,MosMon85b} and bipolar overflows \citep{Puz85}. 
Following this early evolutionary phase, the amount of angular momentum stored by a star before it reaches the zero-age main sequence (hereafter ZAMS) can be determined by the gains and losses regulated by the accretion, winds \citep{Scn62,Shu_00}, and locking with the circumstellar environments with magnetic fields \citep{Kol91,Eds_93}. The shaping of the internal distribution of the stored angular momentum is certainly a matter of hydrodynamical instabilities \citep{EnlSoa81,Str_84,Pit_90,MarBrr91,Kes_95,Som_93,Som_01}, couplings between radiative and convective regions \citep{Lar_81}, and by the lockings of layers through magnetic fields \citep{Bas03}, whether they be fossil or created by various instabilities \citep{Spt99,Spt02}.

If the quoted processes were to lead to establishing only the rigid rotation profile in stars on the ZAMS, the distributions of angular velocity rates $\Omega/\Omega_{\rm crit}$ ($\Omega_{\rm crit}$ is the critical rotational velocity) would probably be unimodal and Maxwellian \citep{Deh70}. However, according to differences in the initial chemical composition and perhaps other conditions, which may regulate the effectiveness of the processes causing the redistribution of the angular momentum in the stellar interior, multimodal distributions of the $\Omega/\Omega_{\rm crit}$ ratio can be expected. Indeed, bimodal rotational velocity distributions are found by \citet{Gue82} for late B-type stars in clusters. The bimodality is observed among young solar mass stars in Orion \citep{AteHet92,ChiHet96,Het_01,Bas03} and is present also among the A-F stars of the main sequence (hereafter MS), where it is correlated with several spectroscopic characteristics: slow rotation favors the appearance of the Am and Ap phenomena \citep{AbtMol95}. However, the slow rotation of Am stars can be caused by tidal braking because they are known to be binaries \citep{Dei00}, while the Ap characteristics can be favored by the magnetic braking in MS evolutionary phase \citep{Hug_00a,Stn00}.
 
  Using a highly homogeneous set of \ensuremath{v\sin i}\ parameters determined for about 1100 stars, divided into six spectral type groups, carefully cleansed of objects presenting the Am/Ap phenomena and of all known binaries, \citet{Ror_07} showed that late B and early A-type MS stars have genuine bimodal distributions of true equatorial rotational velocities. These authors also noticed a striking lack of slow rotators among the intermediate and late A-type stars. They drew attention to the peculiar behavior of the frequency distribution of the surface angular velocity ratios $\Omega/\Omega_{\rm crit}$ in different spectral type groups. Because these groups are characterized by different average ages, the question is raised whether this peculiarity could be the consequence of some difference in the evolution of the rotation according to the stellar mass.

  It is then tempting to review anew in some detail the velocity distributions of stars in the MS phase studied above using a better resolution in mass and age than in \citet{Ror_07}. This may enable us to detect possible signatures in rotational velocity properties induced by the stellar formation characteristics and/or conditioned by the evolution that follows the ZAMS evolutionary stage.

 The aim of this paper can  be summarized as follows: (i) we intend to estimate fundamental parameters, mainly masses and ages of all stars in the sample, to identify  possible statistical indications on differentiated evolutionary characteristics of the rotational velocity as a function of the stellar mass and age; (ii) to determine what kind of internal rotational distribution may characterize the stars according to their mass and age during their MS phase, which may have some effect on the observed surface rotational velocities.

  The selection of the stellar sample studied and the rotational velocity data are described in  Sect.~\ref{rot_data}. The determination of masses and ages is presented in Sect.~\ref{param}. The rotational velocity distributions are presented in Sect.~\ref{dist_vel}, with details on the statistical processing of stellar samples to obtain the equatorial velocity distributions from the observed \ensuremath{v\sin i}\ values. The evolution of these distributions with age is also discussed. Section\,\ref{evol_rot_vel} compares the observed evolution of rotational velocities with different theoretical models. Finally, the results are summarized and discussed in Sect.~\ref{disc_concl}.


\begin{figure*}[!htp]
\centering
\resizebox{\hsize}{!}{\includegraphics{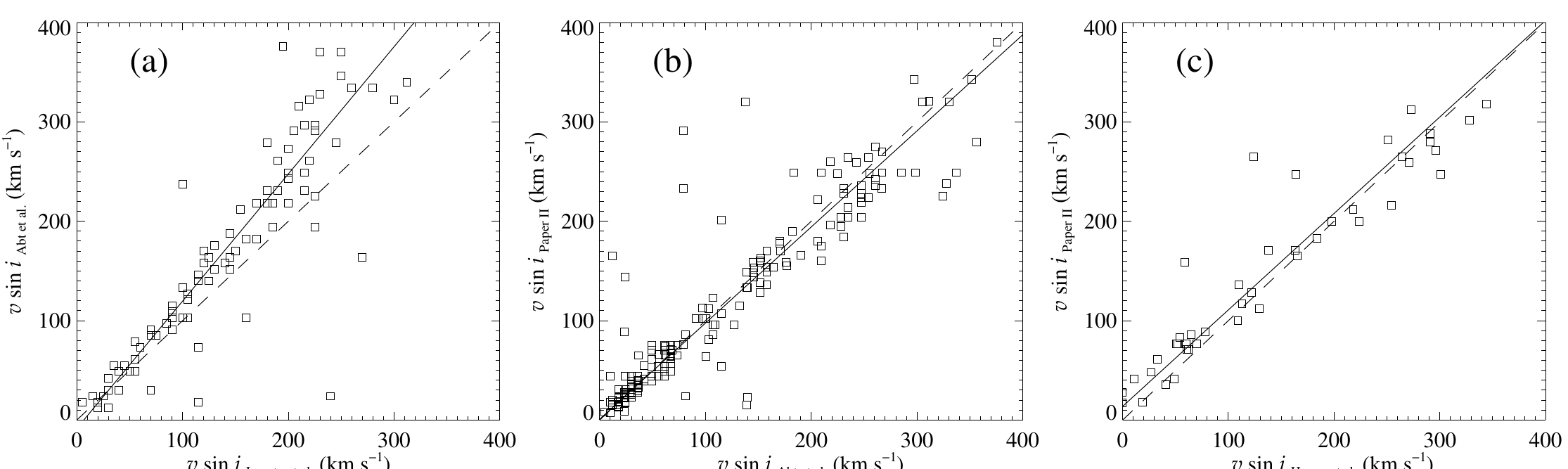}}
     \caption{Comparison of \ensuremath{v\sin i}\ data: (a) Values for the 94 stars (with spectral type equal to, or later than, B6) in common between \citet{Abt_02} and \citet{LeoGro04}. The dashed line stands for the one-to-one relation and the solid line is the result from Eq.\,\ref{abt_leo}. (b) Values for the 199 stars (with spectral type equal to, or later than, B6) in common between merged data from \citet{Abt_02} and \citet{LeoGro04}, and \citet{Ror_02b}. The solid line is the result from Eq.\,\ref{ourwork_abtleo}. (c) Values for the 44 stars (with spectral type equal to, or later than, B6) in common between \citet{Hug_10} and previously merged data. The solid line is the result from Eq.\,\ref{ourwork_huang}.}
     \label{vsinicomp}
\end{figure*}

\section{Rotational velocity data}

  The sample of stars used in the present work is primarily determined by the objects studied in \citet[hereafter Papers I, II, and III]{Ror_02a,Ror_02b,Ror_07}. Because we attempt to study the rotational characteristics only of A-type stars in the MS evolutionary phase, the basis of our sample is the selection of dwarf stars with luminosity class from V to IV made in Paper~III. However, the selection on the spectral class, from B9 to F2, is similar to a cut in temperature and strongly affects the distribution in the  plane mass--age. The first half of the MS, corresponding to young stars, is not sampled in the hot end of the selection (spectral types B9 to A1). In order to cover the entire age sequence for stars more massive than about $2.5\,\ensuremath{M_\odot}$ that correspond to early A and late B stars, we chose to extend the spectral type range up to B6. 

The final selection of objects that comply with the purpose of the present study, however, was made using the masses and ages derived using the effective temperatures and bolometric luminosities determined in the present contribution as detailed in Sect.~\ref{mass_ages}.We briefly recall the main characteristics of the basic sample gathered in the above cited papers in the following subsections.
 
\label{rot_data}
\subsection{\ensuremath{v\sin i}\ data sources}

Our study is based on four different sources of \ensuremath{v\sin i}\ data, from which we collected both B- and A-type stars: 
\begin{itemize}
\item Paper~II provides a large and homogeneous \ensuremath{v\sin i}\ data set for A-type stars by  merging data from \citet{AbtMol95} with accurate \ensuremath{v\sin i}\ determined by Fourier transforms (FT). The first set is scaled to the later.
\item \citet[hereafter called ALG]{Abt_02} provide homogeneous \ensuremath{v\sin i}\ for 1092 B-type stars derived from the FWHM of \ion{He}{i} 4471 and \ion{Mg}{ii} 4481\,\AA\ lines calibrated into the \ensuremath{v\sin i}\ scale of \citet{Slk_75}. This is the extension of the work from \citet{AbtMol95} toward B-stars and it is limited to stars brighter than $V=6.5$\,mag.
\item \citet[hereafter called LG]{LeoGro04} give the southern counterpart of ALG's work: \ensuremath{v\sin i}\ for 1027 B-type stars determined with the same method. 
\item \citet[hereafter called HGM]{Hug_10} study rotation in a sample of B-stars (in open clusters and field). Their \ensuremath{v\sin i}\ determination was made by fitting synthetic model profiles of \ion{He}{i} 4471 and \ion{Mg}{ii} 4481\,\AA\ lines.
 Only their field star data are taken into account in this work.
\end{itemize}

This study covers the spectral types from B6 to F2. The number of stars per spectral type is given in Table\,\ref{mass}.
In the total list of our sample stars (Table\,\ref{fund_par}) we provide the source of the \ensuremath{v\sin i}\ data: sources 1, 2, and 4 (and their combinations) are taken from Paper~II, source 8 is taken from ALG, 16 from LG and 32 from HGM.

\subsection{Merging and scaling \ensuremath{v\sin i}\ data}

Although in Paper~III a merging with data from ALG was already performed to increase the number of stars corresponding to the spectral types B9 and B9.5, a new merging was performed in this work on a wider range of spectral types. 

The data sets from ALG and LG were first merged, considering only stars later than B6. Then the merged sample ALG$\cup$LG was merged with the A-type star sample from Paper~II. Finally \ensuremath{v\sin i}\ of field B-type stars from HGM were merged with the full sample.
The different \ensuremath{v\sin i}\ scales are compared in Fig.~\ref{vsinicomp} for the 94 stars in common between ALG and LG, and for the 199 stars in common between Paper~II and ALG$\cup$LG, and for the 44 stars in common between Paper~II$\cup$ALG$\cup$LG and HGM. 

The linear regression lines between the different scales are determined with GaussFit \citep{Jes_98a,Jes_98b}, a robust least-squares minimization program, to obtain empirical functions:
\begin{equation}
\ensuremath{v\sin i}_\mathrm{ALG} = 1.28{\scriptstyle\pm 0.03}\,\ensuremath{v\sin i}_\mathrm{LG} -8.5{\scriptstyle\pm 2.5}.
\label{abt_leo}
\end{equation}

\begin{equation}
\ensuremath{v\sin i}_\mathrm{Paper II} = 0.967{\scriptstyle\pm 0.19}\,\ensuremath{v\sin i}_\mathrm{ALG} + 0.84{\scriptstyle\pm 0.9}.
\label{ourwork_abtleo}
\end{equation}

\begin{equation}
\ensuremath{v\sin i}_\mathrm{Paper II} = 0.971{\scriptstyle\pm 0.04}\,\ensuremath{v\sin i}_\mathrm{HGM} + 13.7{\scriptstyle\pm 5.1}.
\label{ourwork_huang}
\end{equation}
A 10\% error on the \ensuremath{v\sin i}\ values was assumed for all data when using GaussFit to derive these relations.

 Although the \ensuremath{v\sin i}\ from \citet{Slk_75} are affected by the gravitational darkening effect, \citet{Hoh04} showed that they are underestimated compared with more recent gravitational darkening-dependent determinations of \ensuremath{v\sin i}\ \citep{Frt_05a}. 
 Except for the assumptions that underlie the FT method, the \ensuremath{v\sin i}\ scale defined in Paper~II is independent of any other calibration. This scale is moreover confirmed by a more robust FT method \citep{ResRor04a} and proves to be consistent also with the \ensuremath{v\sin i}\ determinations by \citet{ErrNoh03} and \citet{Fel03}. Recent results from \citet{Diz_10} provide accurate measurements of \ensuremath{v\sin i}\ using FT of the cross-correlation function for 251 A-type stars in the southern hemisphere. Their method is much less sensitive to blends than the individual line measurements made in Paper~I, and their \ensuremath{v\sin i}\ values are 5\% higher for $\ensuremath{v\sin i}>150$\,\ensuremath{\mbox{km}\,\mbox{s}^{-1}}. This effect is smaller than the 10\% error assumed for our merged sample, however, and because of the late knowledge of their results, we chose to leave our scale unchanged.
We therefore used the FT scale of \ensuremath{v\sin i}\ defined in Paper~II and used Eqs.\,\ref{abt_leo}, \ref{ourwork_abtleo}, and \ref{ourwork_huang} to scale all \ensuremath{v\sin i}\ data to this scale. The full sample contains 2014 stars and the homogenized \ensuremath{v\sin i}\ are given in Table\,\ref{fund_par}.

\subsection{Chemically peculiar and binary stars}

  In order to have a stellar sample that is as clean as possible from objects whose rotation could have been modified by tidal or magnetic braking, all known chemically peculiar stars (CP) and ``close'' binary stars (CB) were discarded. The selection was made with the same criteria as we used in previous papers.

\emph{CP stars} (chemically peculiar stars): the catalog of Ap and
Am stars from \citet{RenMad09} and the spectral classifications given by ALG, LG and HGM were used to identify peculiar stars.

\emph{CB stars} (``close'' binary stars): This category of stars was selected on criteria based on HIPPARCOS and spectroscopic data. Most of the selected stars are in the HIPPARCOS catalog \citep{Hip}. The binaries detected by the satellite with $\Delta Hp < 4$ mag are flagged as CB stars. The Ninth Catalog of Spectroscopic Binary Orbits \citep{Pox_04} was used to complete the identification.

All stars that do not obey CB or CP criteria are simply called ``normal".
The question of identification completeness is studied in Paper~II, which concludes that the fraction of CP stars to all stars in the merged sample is fairly constant for all magnitudes brighter than $V = 6.5$ mag and represents roughly 15\% of the objects.

\section{Fundamental parameters}
\label{param}

\begin{table*}[!htp]
\caption{\textbf{(extract)} List of the 2014 B6- to F2-type stars, with their spectral type and fundamental parameters.}
\label{fund_par}
\centering
\begin{tabular}{rlccccccccccc}
\hline
\hline	
HD & Spec. type & $\log\ensuremath{T_\mathrm{eff}}$ & $\sigma_T$ & $\log L/L_\odot$ & $\sigma_L$ & $M/\ensuremath{M_\odot}$ & $\sigma_M$ & $\ensuremath{t/t_\mathrm{MS}}$ &  $\sigma_{\ensuremath{t/t_\mathrm{MS}}}$ & \ensuremath{v\sin i} & Source & Classification \\
 & & & & & & & (\ensuremath{M_\odot}) & & & (\ensuremath{\mbox{km}\,\mbox{s}^{-1}}) & & \\
\hline
  3   & A1Vn   & 3.957 & 0.008 & 1.6387 & 0.0725 & 2.36 & 0.08 & 0.685 & 0.075 & 228 &  4 & CP \\    %
  203 & F2IV   & 3.832 & 0.003 & 0.6669 & 0.0146 & 1.42 & 0.01 & 0.449 & 0.048 & 170 &  4 &    \\    %
  256 & A2IV/V & 3.943 & 0.009 & 2.0664 & 0.1160 & 2.79 & 0.12 & 0.983 & 0.040 & 241 &  5 &    \\    %
  319 & A1V    & 3.945 & 0.009 & 1.4761 & 0.0306 & 2.18 & 0.03 & 0.592 & 0.062 &  59 &  5 & CP \\    %
  431 & A7IV   & 3.889 & 0.011 & 1.5825 & 0.0554 & 2.23 & 0.06 & 0.899 & 0.042 &  97 &  4 & CB \\    %
  560 & B9V    & 4.035 & 0.004 & 1.8414 & 0.0293 & 2.74 & 0.03 & 0.417 & 0.054 & 249 &  1 &    \\    %
  565 & A6V    & 3.906 & 0.004 & 1.6779 & 0.0422 & 2.35 & 0.05 & 0.900 & 0.025 & 149 &  1 &    \\    %
  584 & B7IV   & 4.116 & 0.003 & 2.1390 & 0.0740 & 3.38 & 0.10 & 0.140 & 0.127 &  12 &  8 &    \\    %
  709 & B8     & 4.078 & 0.002 & 3.2140 & 0.5320 & 5.17 & 1.60 & 0.988 & 0.060 & 264 & 32 & CB \\    %
  ... \\
\hline
\end{tabular}
\tablefoot{$\sigma_T$ and  $\sigma_L$ are the errors in logarithmic \ensuremath{T_\mathrm{eff}}\ and logarithmic luminosity. The homogenized \ensuremath{v\sin i}\ is given (with corresponding source) as well as the classification (CP, CB, blank stands for normal). The source flag is the same as in Paper~II, and additionally 8 for ALG, 16 for LG, and 32 for HGM.}
\end{table*}

  In the present contribution, the rotational properties of stars are studied as a function of age and mass following a statistical approach. The studied objects were gathered into groups that contain enough data to warrant that the projection effect in \ensuremath{v\sin i}\ can be statistically corrected. The subtle deviations caused by fast rotation effects, in particular those affecting the \ensuremath{v\sin i}\ parameter \citep{Zoc_02,Frt_05a}, can easily be obliterated in the averaging process by the uncertainties affecting the individual fundamental parameters. Therefore we inferred stellar masses and ages from the apparent effective temperatures and bolometric luminosities in the present approach as if the objects were at rest.

\subsection{Effective temperatures and bolometric luminosities}
\label{eff_lum}

\begin{figure}[!htp]
\centering
\resizebox{\hsize}{!}{\includegraphics{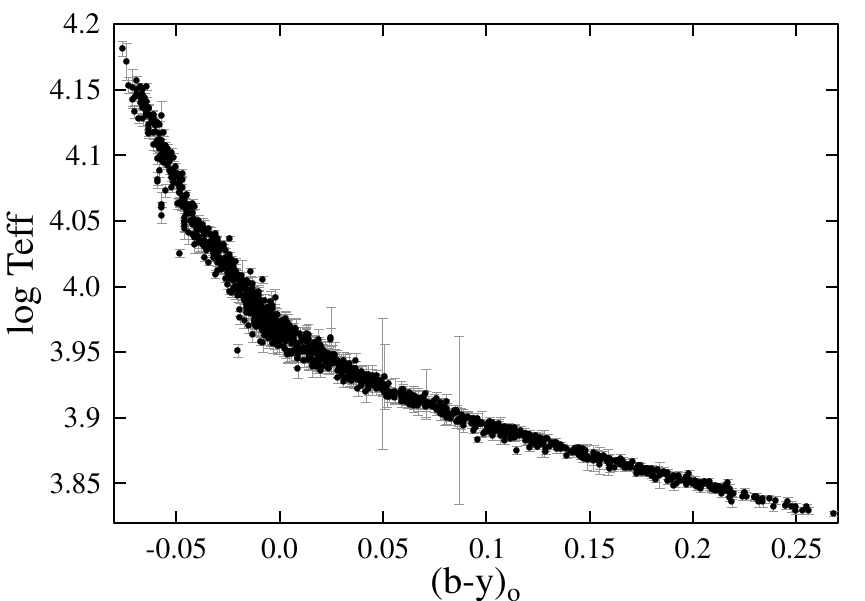}}
\caption{Color-effective temperature diagram for the single normal stars
selected for the present study.}       
\label{teff_by}
\resizebox{\hsize}{!}{\includegraphics{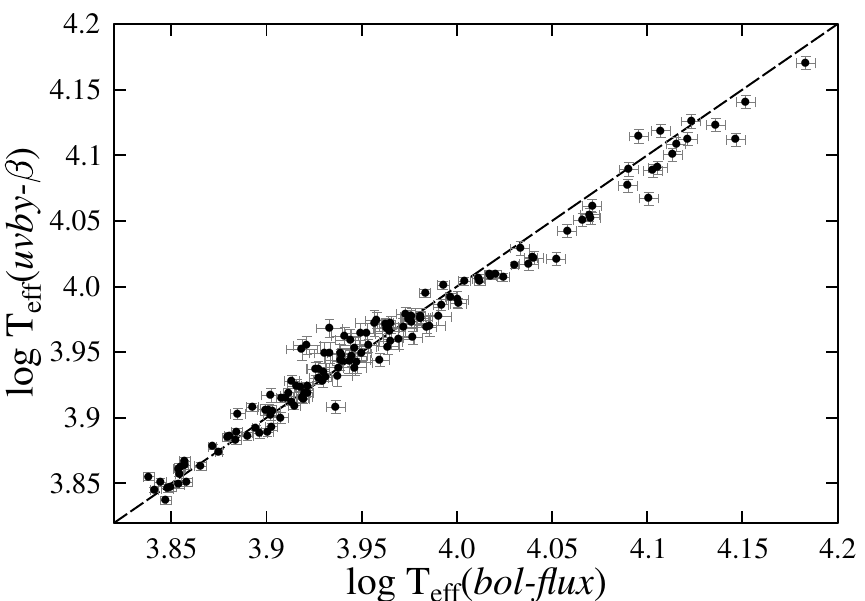}}
\caption{Comparison of effective temperatures calculated in this paper from the $uvby-\beta$ photometry and those for 135 stars obtained using bolometric fluxes. The dashed line is the one-to-one relation.}       
\label{teff_bol}
\end{figure}

   The effective temperatures of stars were derived using $uvby$--$\beta$ Str\"omgren photometry and the calibration of Str\"omgren photometric color indices into \ensuremath{T_\mathrm{eff}}. The calibrations employed are from \citet{MonDwy85}, its revisited version for some spectral types and luminosity classes obtained by \citet{Cai91}, and the corrected and extended one for the hottest stars by \citet{Nai_93}. The adopted \ensuremath{T_\mathrm{eff}}\ for a given star is the average obtained as follows. The $(b-y)$, $m_1$, $c_1$ and $\beta$ indices were taken from \citet{HakMed98} and used with their respective uncertainty. The photometric indices were used following a Monte Carlo sampling within intervals $(\langle X\rangle-2\sigma_X,\langle X\rangle+2\sigma_X$), where $\sigma_X$ is the dispersion of each individual photometric index $X$. For each star we obtained 6500 \ensuremath{T_\mathrm{eff}}-determinations, whose average and the corresponding dispersion are the adopted $\ensuremath{T_\mathrm{eff}}\pm\sigma_{\ensuremath{T_\mathrm{eff}}}$ values presented in Table~\ref{fund_par}. In Fig.~\ref{teff_by} we show the obtained effective temperatures against the photometric colors $(b-y)$ corrected for interstellar extinction. For some stars in the sample of \citet{Hug_10} $uvby$--$\beta$ photometry is not available, so that we used the \ensuremath{T_\mathrm{eff}}\ estimates derived by these authors.

 Sometimes, more or less systematic deviations can be expected between the effective temperatures determined from photometric indices with those based on bolometric fluxes \citep{Zoc_09}. While the details of the stellar energy distribution can be important when \ensuremath{T_\mathrm{eff}}\ is determined with photometric indices, they have a marginal incidence in the \ensuremath{T_\mathrm{eff}}-determination with bolometric fluxes \citep{FrtZoc03,Frt_03}. We compared the effective temperatures of 135 stars obtained in this work with those determined with bolometric fluxes. Techniques for determining effective temperatures with integrated bolometric fluxes are given in \citet{Zoc_09} and they are presently used to determine the effective temperature of 600 A- and F-type stars (in preparation). The comparison (Fig.~\ref{teff_bol}) shows a great consistency.   
 
  The bolometric luminosity of stars is estimated with
\begin{equation}
M_{\rm bol} = M_V+BC(\ensuremath{T_\mathrm{eff}}),
\label{mbol}
\end{equation}
where $M_V$ is the visual absolute magnitude and $BC(T_{\rm eff})$ is the bolometric correction \citep{Lag92}. To calculate $M_V$ we used (i) the apparent visual magnitude of the $UBV$ photometry and the corresponding uncertainty; (ii) the HIPPARCOS parallaxes recently re-determined by \citet{vLe07} and taking into account the respective uncertainties; (iii) the interstellar color excess $E(B-V)$ estimated using the $uvby$--$\beta$ photometry and the calibration of intrinsic colors by \citet{MonDwy85}. Each adopted $M_{\rm bol}$ magnitude is the average of all determinations by Eq.\,\ref{mbol} following a Monte Carlo sampling of the uncertainties affecting \ensuremath{T_\mathrm{eff}}\ and the parameters entering Pogson's relation for $M_V$. The adopted bolometric luminosity parameter $\log L/L_{\odot}$, given in Table~\ref{fund_par} with its $1\sigma$ uncertainty, was derived by adopting $M_{\rm bol}^{\odot}=4.742$ mag.

\subsection{Stellar masses and ages}
\label{mass_ages}
\begin{figure}[!htp]
\resizebox{\hsize}{!}{\includegraphics{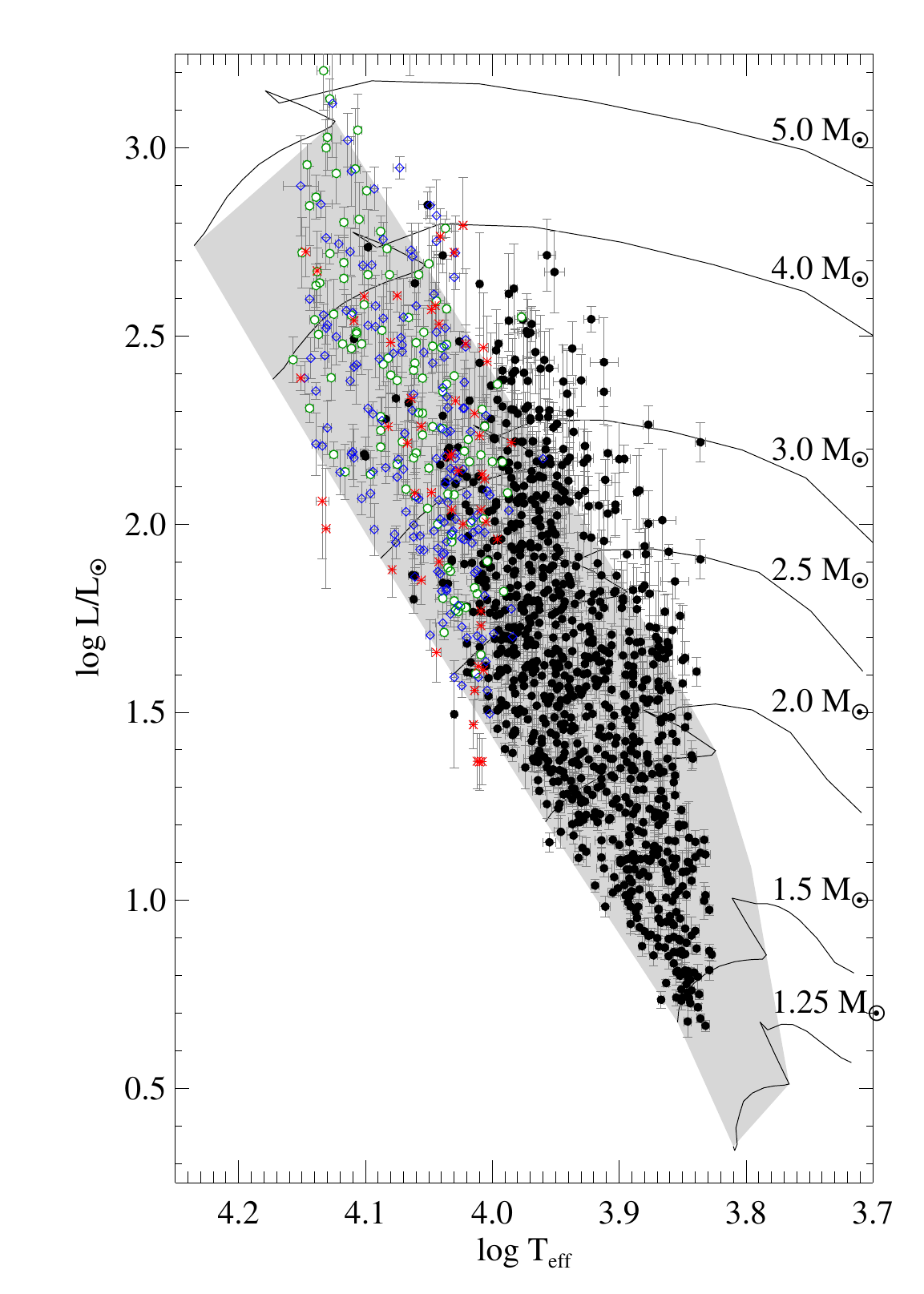}}
\caption{HR diagram of the studied stars. Only the single normal stars in the shaded region were selected for the present study. Filled circles are for stars in the sample studied by \citet{Ror_07}. Open circles, open diamonds and asterisks stand for complementary stars taken from \citet{Abt_02}, \citet{LeoGro04} and \citet{Hug_10} respectively. The evolutionary tracks are from \citet{Scr_92}.}        
\label{hr_diag}
\end{figure}

\begin{figure}[!htp]
\centering
\resizebox{\hsize}{!}{\includegraphics{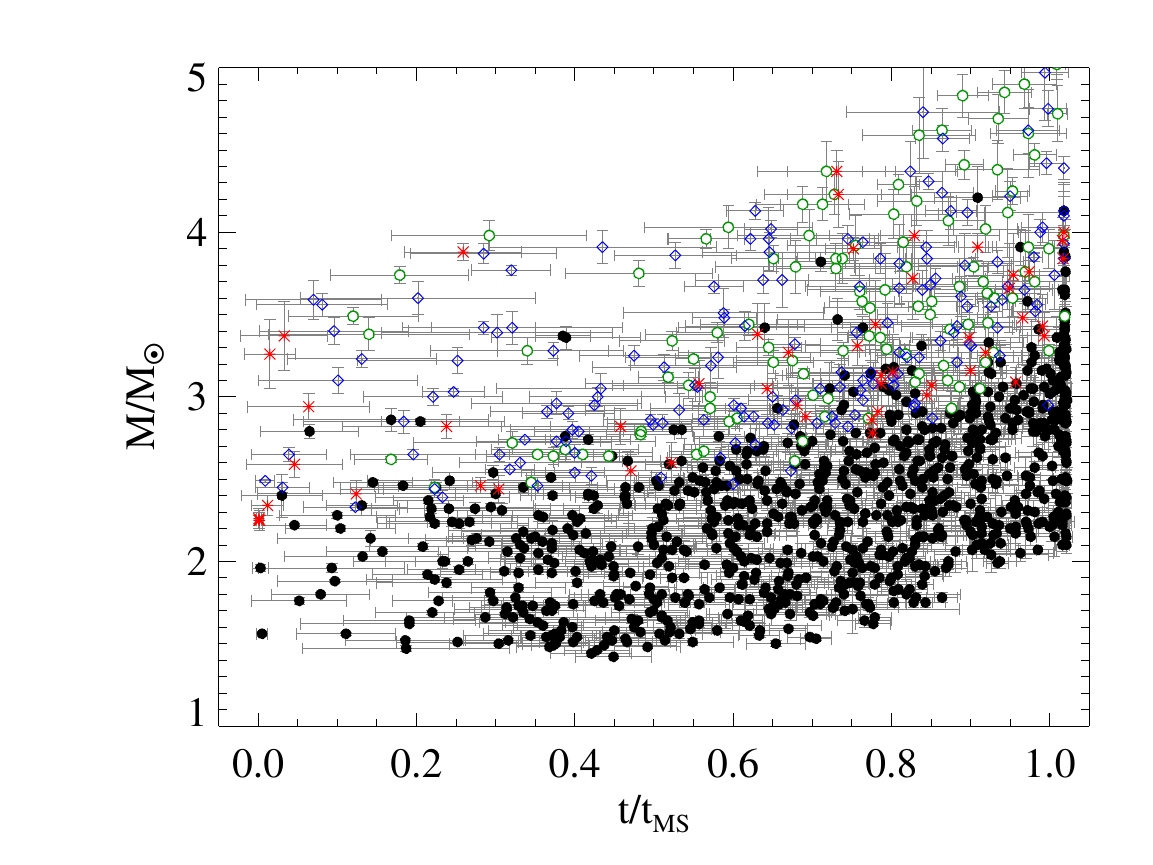}}
\caption{Distribution in the $(\ensuremath{t/t_\mathrm{MS}},M/\ensuremath{M_\odot})$ plane of the data concerning the single normal stars selected for the present study. The error bars in both coordinates are indicated in gray. The symbols have the same meaning as in Fig.\,\ref{hr_diag}.}       
\label{mass_age}
\end{figure}

  To infer the individual stellar masses and ages, we used the pairs $(\ensuremath{T_\mathrm{eff}}, \log L/L_{\odot})$ obtained above with their respective $2\sigma$ deviations as entries in the evolutionary diagrams of non-rotating stars calculated by \citet{Scr_92} that we display in Fig.\,\ref{hr_diag}. The interpolation of masses and ages also proceeds with a Monte Carlo sampling of the $(\ensuremath{T_\mathrm{eff}}, \log L/L_{\odot})$ entries. Table~\ref{fund_par} gives the average $(M/\ensuremath{M_\odot},\ensuremath{t/t_\mathrm{MS}})$, i.e. mass and fractional age parameters with the $1\sigma$ dispersion that are the results from each interpolation procedure. We use here the notation $t$ for the age in years, and \ensuremath{t/t_\mathrm{MS}}\ is the life span from the ZAMS to the TAMS (terminal age main sequence) of a star with the given mass $M/\ensuremath{M_\odot}$. Figure\,\ref{mass_age} shows the distribution in the mass--age diagram of the MS stars used in the present study. The vertical cuts in effective temperature in Fig.\,\ref{hr_diag} that appear because of the selection based on MK spectral type are reflected in Fig.\,\ref{mass_age} by the limits in mass varying with age. Table\,\ref{mass} gives some statistical estimators that describe the characteristics of the stellar mass distribution tabulated against the spectral types of the present analysis.

The stars with large error bars on mass and/or age ($\sigma/\ensuremath{M_\odot}\ge0.3$, $\sigma_{\ensuremath{t/t_\mathrm{MS}}}\ge 0.15$) were discarded for the next steps of this work.

\begin{table}[!htp]
\caption[]{Parameters of the mass distributions in different spectral type intervals.}
\centering
\begin{tabular}{lcccccc}
\hline
\hline	
Spectral  &  \#   &   $M_1$   & $\sigma_1$ & $M_2$ & $\sigma_2$ & $M_3$ \\
type      & stars &  (\ensuremath{M_\odot})   &(\ensuremath{M_\odot}) &(\ensuremath{M_\odot})  &(\ensuremath{M_\odot})  & (\ensuremath{M_\odot}) \\
\hline	
B6--B7&     135&4.02&0.51&4.24&0.64&4.09\\
B8&         182&3.43&0.54&3.52&0.54&3.44\\
B9--B9.5&   328&2.99&0.45&3.08&0.55&2.99\\
A0&         193&2.57&0.36&2.66&0.42&2.59\\
A1&         193&2.51&0.38&2.59&0.35&2.53\\
A2&         194&2.47&0.43&2.50&0.41&2.50\\
A3&         170&2.27&0.38&2.39&0.45&2.30\\
A4--A6&     179&2.11&0.31&2.24&0.43&2.14\\
A7--A9&     180&1.93&0.31&1.99&0.49&1.93\\
F0--F2&     194&1.68&0.26&1.84&0.39&1.74\\
\hline
\end{tabular}
\tablefoot{The values $M_1$ and $\sigma_1$ are the center and sigma of the Gaussian fit of the distribution, $M_2$ and $\sigma_2$ are the mean and the standard deviation of the distribution and $M_3$ its median value.}
\label{mass}
\end{table}

\subsubsection{Uncertainties regarding the stellar masses and ages}
\label{curma}

\begin{table}[!ht]
\caption[]{Mass and age estimates using evolutionary tracks with rotation for a test star with apparent rotation-less mass $M=3\,\ensuremath{M_\odot}$.}
\label{mt_omega}
\centering
\begin{tabular}{c|ccc|ccc}
\hline\hline	
&\multicolumn{3}{c}{$M(\Omega)/\ensuremath{M_\odot}$} & \multicolumn{3}{|c}{$t(\Omega)/t_{\rm MS}(\Omega)$} \\
$\Omega/\Omega_{\rm c}$ & $\ensuremath{t/t_\mathrm{MS}}=0.2$ & 0.5 & 0.8 & $\ensuremath{t/t_\mathrm{MS}}=0.2$ & 0.5 & 0.8 \\
\hline	
0.5 & 3.05 & 2.91 & 2.83 & 0.20 & 0.49 & 0.78 \\
0.8 & 3.14 & 3.00 & 2.85 & 0.19 & 0.46 & 0.70 \\
0.9 & 3.18 & 3.07 & 2.89 & 0.19 & 0.44 & 0.67 \\
1.0 & 3.22 & 3.13 & 2.93 & 0.18 & 0.42 & 0.63 \\
\hline
\end{tabular}
\end{table}

Because models of stellar evolution with rotation can produce evolutionary tracks that can differ significantly from those without rotation \citep{MarMet00,Ekm_08,Ekm_11}, we may wonder whether the masses and ages obtained for our sample stars, which can be intrinsic rapid rotators, are still realistic. Because we do not know the actual internal rotation of a given star in advance, whatever uncertainty we can estimate will necessarily be model-dependent. We can then recall that the most frequently used models \citep[c.f.][]{MarMet00,Ekm_08,Ekm_11} are based on two strong assumptions: (i) all stars initiate their evolution in the MS as rigid rotators, which means the total rotational energy stored in the object is strongly limited (see discussion in Sect.~\ref{int_om}); (ii) the redistribution of angular momentum proceeds only over barotropic surfaces and produces only ``shellular'' internal rotation laws.
We have calculated the masses and rotation-dependent masses and fractional ages $t(\Omega)/t_{\rm MS}(\Omega)$ suggested by these models for a test star that according to rotation-less models has a mass $M=3\,\ensuremath{M_\odot}$ and lies at evolutionary phases $\ensuremath{t/t_\mathrm{MS}}=0.2$, 0.5 and 0.8. The obtained results are given in Table~\ref{mt_omega}. From this table we can conclude that the parameters based on rotation-less models used in the present work $(M/\ensuremath{M_\odot},t(\Omega)/\ensuremath{t/t_\mathrm{MS}})$ may slightly underestimate stellar masses in the first half of the MS, and that the differences are larger for a higher ratio $\Omega/\Omega_{\rm crit}$. Masses can be slightly overestimated for stars in the second half of the MS. Because the actual ratio $\Omega/\Omega_{\rm crit}$ is not known, the stellar mass and its uncertainty are on average $\langle M\rangle=3.01\pm0.12\,\ensuremath{M_\odot}$. In Table~\ref{mt_omega} we note that rotation-dependent ratios can be systematically lower than those derived from rotation-less models and the differences are larger the closer the stars are to the TAMS and the higher the ratio $\Omega/\Omega_{\rm crit}$. The effect of neglecting rotation on the derivation of $M/\ensuremath{M_\odot}$ and \ensuremath{t/t_\mathrm{MS}}\ for other stellar masses can be seen in Fig.~6 of \citet{Zoc_05}, where the authors studied Be stars, which are considered as paradigms for rapidly rotating MS stars. We note, however, that the fraction of program stars in the present work with $\Omega/\Omega_{\rm crit}\!\approx\!1.0$ must probably be very low, because the fraction of Be stars among the B-type star population is low \citep{ZocBrt97}. On the other hand, as soon as $\Omega/\Omega_{\rm crit}\!\lesssim\!0.9$, age differences become $\delta(t(\Omega)/t_{\rm MS})\lesssim0.1$, which are on the same order or smaller than the age-bins used to estimate the averages $\langle v\rangle$ of true equatorial velocities in the following sections.

\subsubsection{Determining fundamental parameters with
rotation-dependent evolutionary tracks}
\label{wrumwr}

  The use of evolutionary tracks with rotation implies that we have an estimate of the true equatorial velocity $v$ of the studied star. These tracks are given in terms of bolometric luminosities and effective temperatures averaged over the rotationally deformed object as a function of age and rotational velocities in the ZAMS. Accordingly, the observed apparent aspect angle-dependent effective temperature and bolometric luminosity must be corrected for rotational effects to derive their counterparts for the corresponding rotation-less object \citep{Frt_05a,Zoc_05}. In turn, these must be averaged over a rotationally deformed star, whose degree of geometrical deformation is measured by the ratio $\ensuremath{v/v_\mathrm{crit}}(M,t)$, where $\ensuremath{v_\mathrm{crit}}(M,t)$ is the stellar critical equatorial velocity. Using the curves representing the evolution of the ratio $v(M,t)/\ensuremath{v_\mathrm{crit}}(M,t)$, the stellar velocity ratio in the ZAMS must be identified. Only evolutionary tracks for this specific initial velocity ratio, and for several masses around the sought one, can finally be used to interpolate the mass and age of the studied object. This operation implies a long iteration procedure, where the stellar mass $M/\ensuremath{M_\odot}$, its age \ensuremath{t/t_\mathrm{MS}}\ and the inclination angle $i$ of the rotational axis are iterated simultaneously. Knowing that on the one hand only a small fraction of program stars may be very rapid rotators, and on the other hand this iteration will not bring significant improvements in the estimate of masses and ages of the remaining stars, the approach used in the present work seems justified.

\subsection{The $v\sin i$ parameters and the rapid rotation}
\label{vprr}

 The estimates of \ensuremath{v\sin i}\ parameters can also be affected by the rapid rotation. Rapid rotation induces temperature and gravity inhomogeneities in the stellar surface known as the ``gravity darkening effect''. Owing to the lowering induced on the local effective temperatures in the equator, these regions do not contribute efficiently to the Doppler rotational broadening of spectral lines. The apparent \ensuremath{v\sin i}\ of rapid rotators must then be corrected for this underestimation \citep{Frt_05a}. This correction can increase our estimates of \ensuremath{v/v_\mathrm{crit}}\ only if it can be demonstrated that in each bin used to average the apparent \ensuremath{v\sin i}\ parameters, the number of rapid rotators with actual ratios $\Omega/\Omega_{\rm c}\gtrsim0.8$ represents a significant fraction of objects.

\section{Distributions of rotational velocities}
\label{dist_vel}
\begin{figure*}[!htp]
\resizebox{\hsize}{!}{\includegraphics{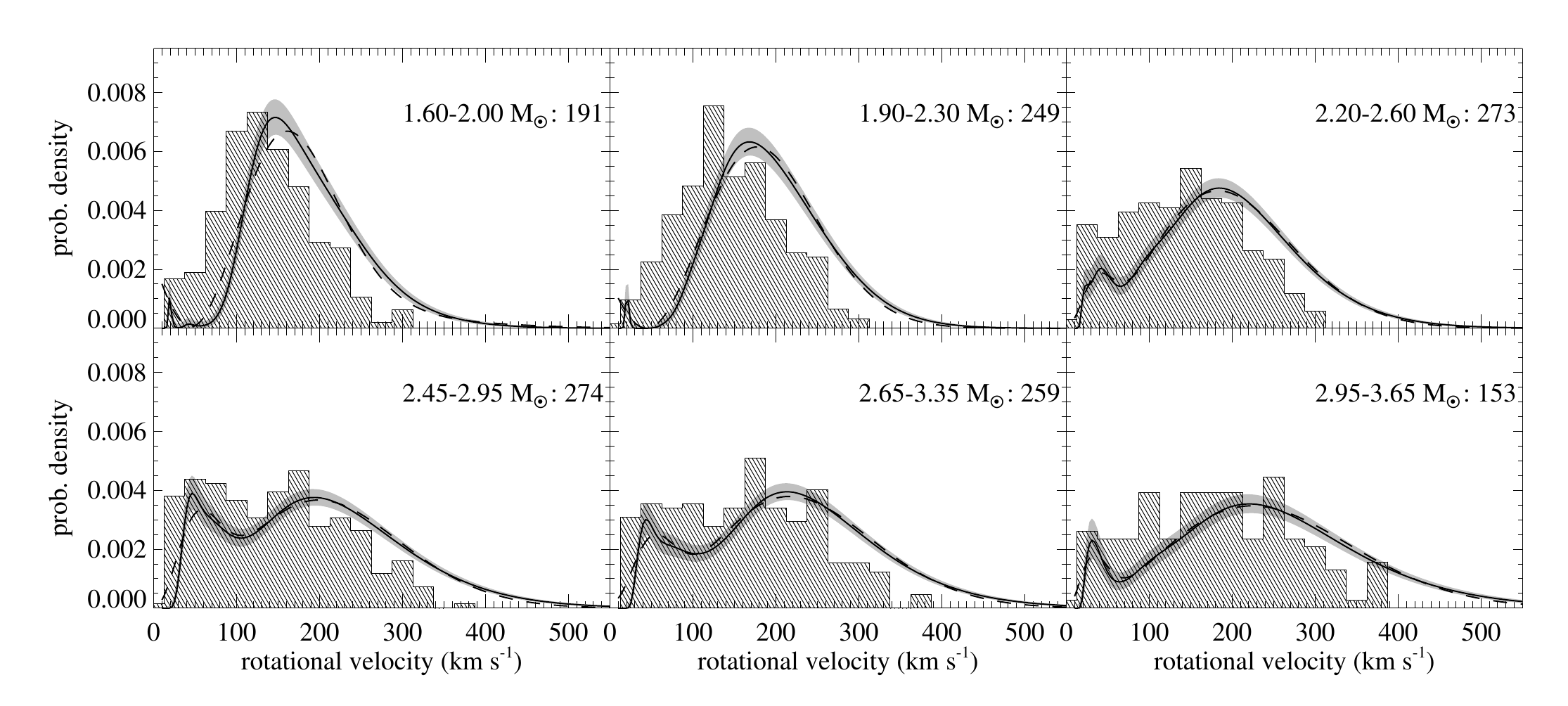}}
     \caption{Distributions of rotational velocities for normal stars in different mass subsamples: shaded histograms are the observed \ensuremath{v\sin i}; solid thick lines are the distributions of equatorial velocities and the gray strips are their associated variability bands. The dashed lines stand for the Maxwellian fit, whose parameters are given in Table~\ref{per_mode}. The mass range for each subsample is indicated in the corresponding panel, together with the number of stars. The histograms are normalized to fit the probability density scale.}
     \label{distrib2}
\end{figure*}

\begin{figure*}[!htp]
\centering
\resizebox{\hsize}{!}{\includegraphics{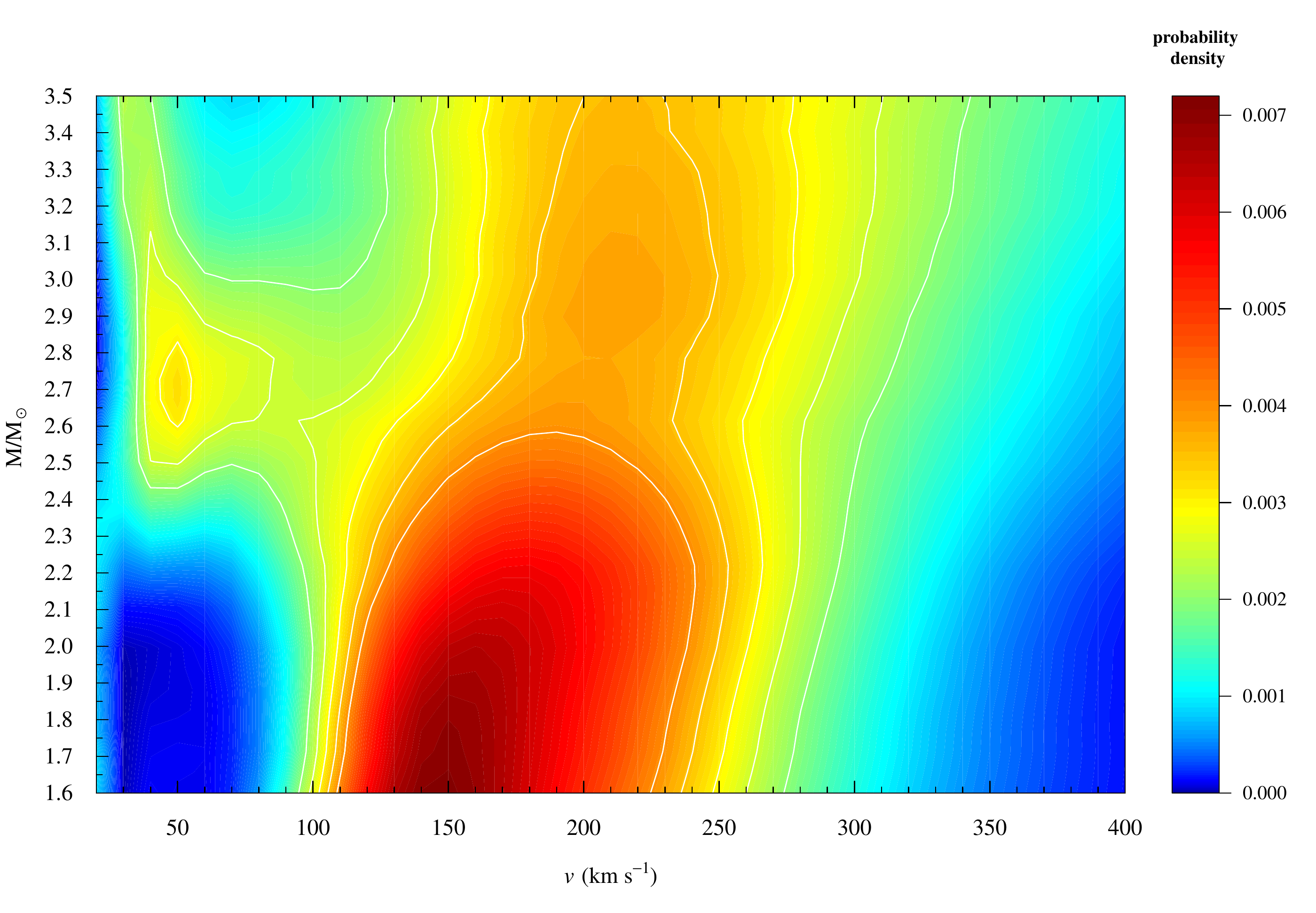}}
\caption{Distribution of true rotational velocities \ensuremath{v}\ for normal stars as a function of the stellar mass 
$M/\ensuremath{M_\odot}$. The color scale represents the density in the one-dimensional normalized distributions. The bluer the region, the smaller the number of stars and vice versa for the red colored scale. White solid lines follow iso-density contours.}       
\label{over_rot}
\end{figure*}

 In Paper~III, distributions of rotational velocities were obtained by dividing the stellar sample into groups of spectral types and mixing stars of all ages in the MS life span. However, the spectral type of a star changes during the MS life span, so that in a given group objects with different masses and ages can display roughly the same apparent spectral type. 
 To have a better representation of the percentage of stars as a function of the rotational velocity per mass-groups and study the distribution of rotational velocities as a function of the stellar mass, the sample of normal stars defined above was divided into overlaping mass intervals that run from 1.6 to 3.85\,\ensuremath{M_\odot}. The limits of the intervals together with the corresponding number of stars are given in Table\,\ref{per_mode}. 
 For each subset, the distribution of \ensuremath{v\sin i}\ was processed in a way similar as in Paper~III:
 \begin{itemize}
 \item the histogram was smoothed using a Gaussian kernel estimator and smoothing parameter $\hat{h}$ defined by \citet{BonAzi97}\footnote{This smoothing parameter is larger than the estimation using the definition by \citet{ShrJos91}, made in Paper~III. This is the reason why the distributions are smoother \citep{Sin86}. This smoothing parameter has been used in the different parts of the study, both in one and two dimensions (see Sect.\,\ref{obs_vel_evol}).}, and variability bands were estimated \citep{BonAzi97},
 \item the smoothed distribution was rectified from the projection effect, assuming randomly oriented rotation axes, with the Lucy-Richardson method \citep{Luy74,Rin72},
 \item the rectified distributions were fitted by a sum of two Maxwellians to derive the position of the mode(s) and the proportion of fast and slow rotators if a significant bimodality is present. The Maxwellian distribution used to fit the fast part of the velocity distribution has an additional lag parameter $\ell$, to be able to shift it toward higher velocities\footnote{\label{lagmaxwell}The lag parameter $\ell$ allows us to shift the Maxwellian distribution along the $x$-axis: $f(x)=(x-\ell)^2\,\sqrt{2/\pi}/\alpha^3\,\exp\left(-\left(x-\ell\right)^2/\left(2\alpha^2\right)\right)$.}.
 \end{itemize}
 \begin{table*}[!htp]
\caption{Parameters of the 1D velocity distributions for normal stars as a function of the stellar mass.} 
\centering
\begin{tabular}{rcc|cccc|cccccc}
\hline\hline
\multicolumn{1}{c}{Mass} & \#   &     & \multicolumn{4}{c|}{Slow rotators} & \multicolumn{6}{c}{Fast rotators} \\
\multicolumn{1}{c}{range}& stars& $\hat{h}$ & \% & $\mu_\mathrm{s}$  & $\mu^\prime_\mathrm{s}$& dispersion & \% & $\sqrt{2}\,\alpha$ & $\ell$ & $\mu_\mathrm{r}$ & $\mu^\prime_\mathrm{r}$& dispersion \\
\multicolumn{1}{c}{(\ensuremath{M_\odot})} & & & & (\ensuremath{\mbox{km}\,\mbox{s}^{-1}})& (\ensuremath{\mbox{km}\,\mbox{s}^{-1}})& (\ensuremath{\mbox{km}\,\mbox{s}^{-1}})& & (\ensuremath{\mbox{km}\,\mbox{s}^{-1}})& (\ensuremath{\mbox{km}\,\mbox{s}^{-1}})& (\ensuremath{\mbox{km}\,\mbox{s}^{-1}})& (\ensuremath{\mbox{km}\,\mbox{s}^{-1}})& (\ensuremath{\mbox{km}\,\mbox{s}^{-1}})\\
\hline
$*$1.60--2.00&191&0.200&  --- & --- &  --- & ---	   &100  & 114 $\pm 1 $ &  46 $\pm 1 $ & 160 &148 & 54 \\
   1.70--2.10&216&0.194&  --- & --- &  --- & ---	   &100  & 124 $\pm 1 $ &  46 $\pm 1 $ & 170 &151 & 59 \\
   1.80--2.20&220&0.177&  --- & --- &  --- & ---	   &100  & 128 $\pm 1 $ &  48 $\pm 1 $ & 176 &160 & 61 \\
$*$1.90--2.30&249&0.190&  --- & --- &  --- & ---	   &100  & 133 $\pm 1 $ &  44 $\pm 1 $ & 177 &169 & 63 \\
   2.00--2.40&276&0.201&  --- & --- &  --- & ---	   &100  & 140 $\pm 1 $ &  45 $\pm 1 $ & 185 &181 & 67 \\
   2.10--2.50&274&0.227&  3   & 25 $\pm 1 $ & 22 & 12 	   & 97  & 154 $\pm 1 $ &  27 $\pm 1 $ & 181 &181 & 73 \\
$*$2.20--2.60&273&0.247&  8   & 37 $\pm 1 $ & 43 & 18 	   & 92  & 164 $\pm 1 $ &  18 $\pm 1 $ & 183 &184 & 78 \\
   2.30--2.70&248&0.260& 12   & 44 $\pm 1 $ & 43 & 21 	   & 88  & 179 $\pm 1 $ &  12 $\pm 2 $ & 191 &187 & 85 \\
   2.35--2.85&282&0.259& 12   & 47 $\pm 1 $ & 49 & 22 	   & 88  & 210 $\pm 1 $ &   0 $\pm 1 $ & 210 &178 &100 \\
$*$2.45--2.95&274&0.254& 18   & 52 $\pm 1 $ & 46 & 25 	   & 82  & 185 $\pm 3 $ &  14 $\pm 4 $ & 199 &193 & 88 \\
   2.55--3.05&235&0.254& 20   & 61 $\pm 1 $ & 52 & 29 	   & 80  & 191 $\pm 2 $ &  20 $\pm 3 $ & 211 &214 & 91 \\
   2.65--3.15&209&0.253& 15   & 55 $\pm 1 $ & 46 & 26 	   & 85  & 183 $\pm 2 $ &  24 $\pm 3 $ & 208 &205 & 87 \\
$*$2.65--3.35&259&0.249& 14   & 52 $\pm 1 $ & 43 & 25 	   & 86  & 187 $\pm 2 $ &  29 $\pm 3 $ & 216 &214 & 89 \\
   2.75--3.45&225&0.270& 13   & 47 $\pm 1 $ & 40 & 22 	   & 87  & 201 $\pm 2 $ &  18 $\pm 3 $ & 219 &217 & 96 \\
   2.85--3.55&190&0.278& 10   & 39 $\pm 1 $ & 37 & 19 	   & 90  & 202 $\pm 2 $ &  20 $\pm 2 $ & 222 &217 & 96 \\
$*$2.95--3.65&153&0.290&  6   & 32 $\pm 1 $ & 31 & 15 	   & 94  & 224 $\pm 2 $ &   0 $\pm 1 $ & 224 &220 &107 \\
   3.05--3.75&131&0.301&  5   & 31 $\pm 1 $ & 31 & 15 	   & 95  & 229 $\pm 1 $ &   0 $\pm 1 $ & 229 &190 &109 \\
   3.15--3.85&122&0.314&  8   & 32 $\pm 1 $ & 28 & 15 	   & 92  & 224 $\pm 2 $ &  15 $\pm 2 $ & 238 &220 &107 \\
\hline
\end{tabular}
\tablefoot{The defined mass ranges and the corresponding number of normal stars are given. The smoothing parameter $\hat{h}$ is given for each subsample, it is expressed in logarithmic \ensuremath{v\sin i}. For each distribution, the parameters of the Maxwellian fit are listed: percentage of slow and fast velocity distributions, mode and dispersion. For slow rotators, the distribution mode $\mu_\mathrm{s}$ is derived from the Maxwellian fit. The fast rotator distribution is a lagged Maxwellian (see text and footnote\,\ref{lagmaxwell}), and its mode is $\mu_\mathrm{r}\equiv\sqrt{2}\,\alpha+\ell$, where $\alpha$ is the parameter of the Maxwellian distribution. The estimators $\mu^\prime$ are derived as the position of the maximum directly on the 1D distributions. The distributions are shown in Fig.~\protect\ref{distrib2} for six of the eighteen mass ranges, indicated by asterisks.}
\label{per_mode}
\end{table*}
The distributions are shown for some of the mass intervals in Fig.\,\ref{distrib2}, and the results of the fits are given in Table\,\ref{per_mode}.
An overview of the distribution of the rotational velocities as a function of mass for the stellar sample in the present work is displayed in Fig.~\ref{over_rot}. In this figure the full MS life span is considered: $0\le\ensuremath{t/t_\mathrm{MS}}\le1$. The color scale represents the density in the one-dimensional normalized distributions: the bluer the region, the smaller the number of stars and vice versa toward the red scale.

 \begin{table*}[!htp]
\caption{Parameters of the 1D velocity distributions for all stars (normal, CP and CB) as a function of the stellar mass. The columns are the same as in Table\,\ref{per_mode}.}
\centering
\begin{tabular}{ccc|cccc|cccccc}
\hline\hline
Mass & \#   &     & \multicolumn{4}{c|}{Slow rotators} & \multicolumn{6}{c}{Fast rotators} \\
range& stars& $\hat{h}$ & \% & $\mu_\mathrm{s}$  & $\mu^\prime_\mathrm{s}$& dispersion & \% & $\sqrt{2}\,\alpha$ & $\ell$ & $\mu_\mathrm{r}$ & $\mu^\prime_\mathrm{r}$& dispersion \\
(\ensuremath{M_\odot}) & & & & (\ensuremath{\mbox{km}\,\mbox{s}^{-1}})& (\ensuremath{\mbox{km}\,\mbox{s}^{-1}})& (\ensuremath{\mbox{km}\,\mbox{s}^{-1}})& & (\ensuremath{\mbox{km}\,\mbox{s}^{-1}})& (\ensuremath{\mbox{km}\,\mbox{s}^{-1}})& (\ensuremath{\mbox{km}\,\mbox{s}^{-1}})& (\ensuremath{\mbox{km}\,\mbox{s}^{-1}})& (\ensuremath{\mbox{km}\,\mbox{s}^{-1}})\\
\hline
~~~1.60--2.00&283&0.196& --- & --- &  --- & ---      &100 & 122$\pm 1$ &  40$\pm 1$ & 162&145& 58\\
~~~1.70--2.10&321&0.189& --- & --- &  --- & ---      &100 & 127$\pm 1$ &  41$\pm 1$ & 168&151& 60\\
~~~1.80--2.20&342&0.174& --- & --- &  --- & ---      &100 & 132$\pm 1$ &  41$\pm 1$ & 172&163& 63\\
~~~1.90--2.30&385&0.194& --- & --- &  --- & ---      &100 & 142$\pm 1$ &  31$\pm 1$ & 173&175& 67\\
~~~2.00--2.40&429&0.207&  5  & 50 $\pm 2$ & 58 & 24  & 95 & 149$\pm 1$ &  32$\pm 1$ & 181&184& 71\\
~~~2.10--2.50&418&0.225&  4  & 34 $\pm 1$ & 58 & 16  & 96 & 168$\pm 1$ &   9$\pm 1$ & 177&184& 80\\
~~~2.20--2.60&426&0.244& 10  & 39 $\pm 1$ & 37 & 19  & 90 & 176$\pm 1$ &   4$\pm 1$ & 180&184& 84\\
~~~2.30--2.70&391&0.263& 15  & 40 $\pm 1$ & 34 & 19  & 85 & 201$\pm 1$ &   0$\pm 1$ & 201&187& 96\\
~~~2.35--2.85&455&0.261& 19  & 43 $\pm 1$ & 34 & 20  & 81 & 189$\pm 1$ &   0$\pm 1$ & 189&181& 90\\
~~~2.45--2.95&447&0.263& 22  & 44 $\pm 1$ & 37 & 21  & 78 & 192$\pm 1$ &   0$\pm 1$ & 192&190& 92\\
~~~2.55--3.05&385&0.275& 23  & 45 $\pm 1$ & 34 & 22  & 77 & 196$\pm 1$ &   0$\pm 1$ & 196&202& 93\\
~~~2.65--3.15&340&0.275& 20  & 44 $\pm 1$ & 34 & 21  & 80 & 190$\pm 1$ &   0$\pm 1$ & 190&193& 90\\
~~~2.65--3.35&415&0.274& 20  & 42 $\pm 1$ & 31 & 20  & 80 & 203$\pm 1$ &   0$\pm 1$ & 203&205& 97\\
~~~2.75--3.45&352&0.287& 19  & 41 $\pm 1$ & 31 & 20  & 81 & 203$\pm 1$ &   0$\pm 1$ & 203&208& 97\\
~~~2.85--3.55&299&0.301& 16  & 38 $\pm 1$ & 31 & 18  & 84 & 212$\pm 1$ &   0$\pm 1$ & 212&205&101\\
~~~2.95--3.65&233&0.316& 16  & 40 $\pm 1$ & 31 & 19  & 84 & 211$\pm 1$ &   0$\pm 1$ & 211&205&100\\
~~~3.05--3.75&209&0.337& 16  & 36 $\pm 1$ & 31 & 17  & 84 & 224$\pm 1$ &   0$\pm 1$ & 224&178&107\\
~~~3.15--3.85&189&0.359& 18  & 35 $\pm 1$ & 28 & 17  & 82 & 229$\pm 1$ &   0$\pm 1$ & 229&208&109\\
\hline
\end{tabular}
\label{per_mode_all}
\end{table*}

 We clearly notice in Fig.~\ref{over_rot} two main groups of stars as a function of stellar mass with a transition occurring at $M\approx2.5\,\ensuremath{M_\odot}$. In the low-mass part of the diagram, the distribution is unimodal and there is a striking lack of objects in the $0\la \ensuremath{v}\la100$\,\ensuremath{\mbox{km}\,\mbox{s}^{-1}}\ interval of velocities. This gap in the distribution in the plane $(v, M)$ is also noticeable, although less pronounced, in (\ensuremath{v\sin i}, spectral type) as shown in \citet{Ror09}. 
With the aim of testing the effect of our selection of normal stars on the resulting distributions, the same process was applied to all stars (normal, CP and CB). Table\,\ref{per_mode_all} gives the results of the fits for the same mass intervals. The lack is still present even when CP and CB stars are not removed from the sample, although some minor overdensities occur for $\ensuremath{v}\la40$\,\ensuremath{\mbox{km}\,\mbox{s}^{-1}}. 
The effect of discarding CP and CB stars cannot account for the lack of slow rotators, and we conclude that stars with low rotational velocities were not created in the space volume scrutinized with our data.

For stars with $M\ga2.5\,\ensuremath{M_\odot}$, the bimodality of the velocity distribution distinctly appears. The proportion of the slow rotators increases to 20\% around $M\approx2.8\,\ensuremath{M_\odot}$ and decreases toward higher masses. The selection of normal stars removes a large proportion of slow rotators ($0\la \ensuremath{v}\la100\,\ensuremath{\mbox{km}\,\mbox{s}^{-1}}$) in this mass range, as seen when comparing the percentage of the slow rotator Maxwellian distributions from Tables\,\ref{per_mode} and \ref{per_mode_all}, but the bimodality remains significant.

The fast rotator mode of the equatorial velocity distribution for normal stars, $\mu_\mathrm{r}$, behaves monotonically over the full range of mass and smoothly increases with stellar mass. This variation can be fitted by a linear relation in the mass range 1.6--3.5\,\ensuremath{M_\odot}:
\begin{equation}  
\mu_\mathrm{r} \approx 41 {\scriptstyle\pm 1.9}\,M/\ensuremath{M_\odot} + 90{\scriptstyle\pm 5}\;\ensuremath{\mbox{km}\,\mbox{s}^{-1}}.
\label{fastmode}
\end{equation}

\subsection{Normal versus peculiar stellar populations}
\label{normal_CP}
It was argued by \citet{Abt09} that slowly rotating A0--A3 ``normal'' stars are stars that did not have enough time to become Ap or Am stars yet, and that Ap(SrCrEu) stars need about half of their MS lifetime to show their chemical peculiarity. This effect would bias the distributions of rotational velocities when one considers only ``normal'' stars and applies a selection by age.
\begin{figure}[!htb]
\centering
\resizebox{\hsize}{!}{\includegraphics{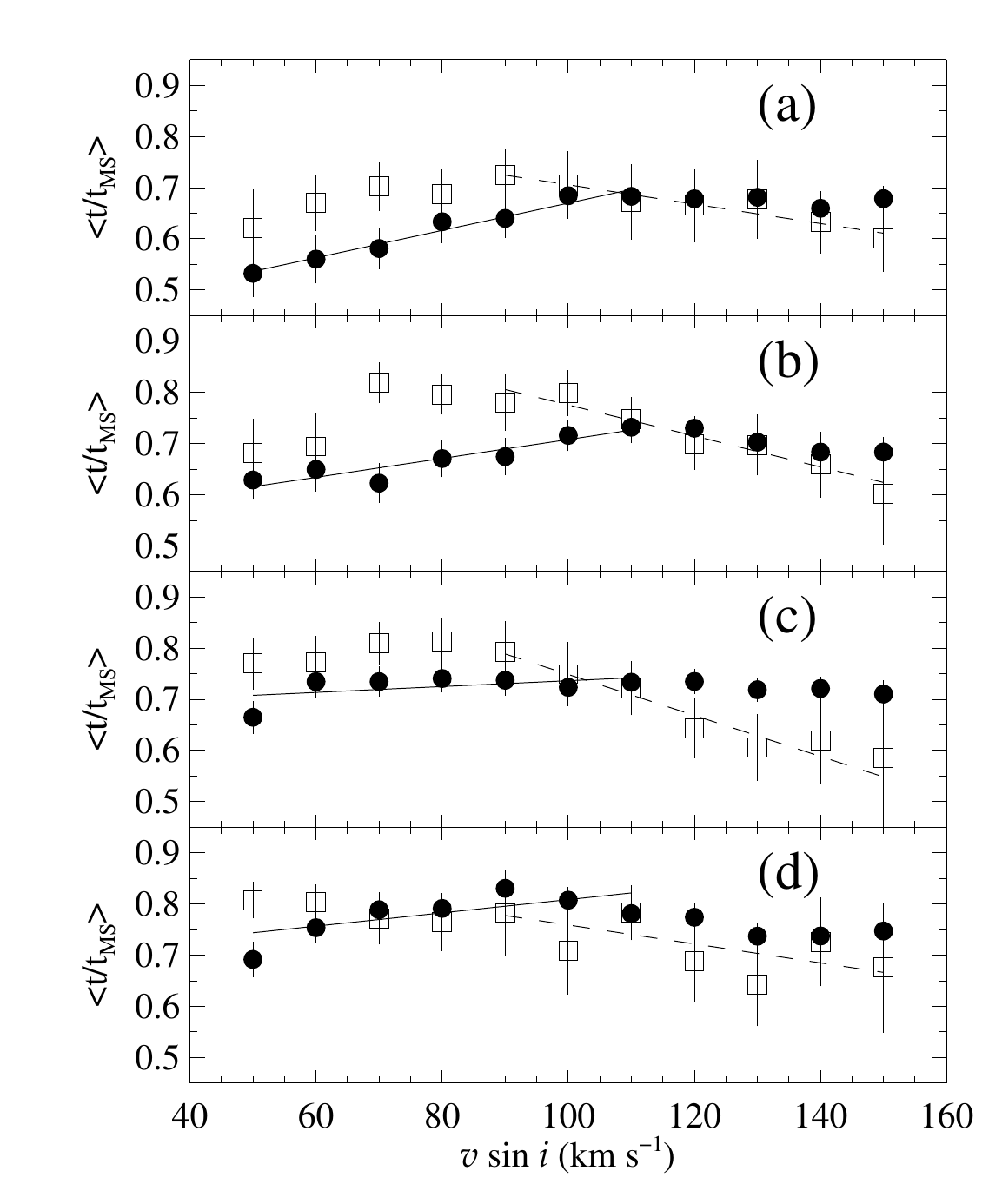}}
\caption{Mean age $\langle t/t_{\rm MS}\rangle$ per bin of \ensuremath{v\sin i}\ for different mass intervals (taken from Table\,\ref{per_mode}): (a) $2<M/\ensuremath{M_\odot} <2.4$; (b) $2.2<M/\ensuremath{M_\odot} <2.6$; (c) $2.35<M/\ensuremath{M_\odot} <2.85$; (d) $2.55<M/\ensuremath{M_\odot} <3.05$. Filled circles stand for normal stars, whereas open squares represent CP stars. The bins in \ensuremath{v\sin i}\ are 40\,\ensuremath{\mbox{km}\,\mbox{s}^{-1}}\ wide. Solid lines are the results of linear fits on normal stars with $\ensuremath{v\sin i}\le110$\,\ensuremath{\mbox{km}\,\mbox{s}^{-1}}\ and dashed lines the results of the fits for CP stars with $\ensuremath{v\sin i}\ge90$\,\ensuremath{\mbox{km}\,\mbox{s}^{-1}}.
}
\label{age-pec}       
\end{figure}

The averaged age on the MS as a function of \ensuremath{v\sin i}\ was derived for different mass intervals in our sample, considering normal stars and CP stars separately (Fig.\,\ref{age-pec}). In the first two mass intervals ($2<M/\ensuremath{M_\odot} <2.6$) there is a significant and systematic shift between normal and CP stars for $\ensuremath{v\sin i}\lesssim 100$\,\ensuremath{\mbox{km}\,\mbox{s}^{-1}}, CP being on average older by 10\% of $t_{\rm MS}$. One should bear in mind that when deriving the fundamental parameters of our targets, normal and CP stars are treated in the same way. Therefore, the bolometric luminosity of CP stars can be overestimated. \citet{Lat_07} indeed show that the bolometric correction for Ap stars is smaller than the one calibrated by \citet{LazCaa92} by an amount of about 0.2\,mag. Their age derived from their position in the HR diagram may consequently be also overestimated. 

On the other hand, a trend is seen in the variation of the mean age of normal stars in the first mass interval ($2<M/\ensuremath{M_\odot} <2.4$). It increases monotonically from 53\% to 68\% of $t_{\rm MS}$ between $\ensuremath{v\sin i}=50$ and 110\,\ensuremath{\mbox{km}\,\mbox{s}^{-1}}, and remains constant for faster rotators. This trend is less pronounced in the next mass bin, and completely disappears for more massive stars.
This trend agrees with that postulated by \citet{Abt09}. Moreover, the variation of the mean age with rotation is recovered using \ensuremath{v\sin i}, therefore the projection effect can lessen the slope. But the selection here was made using mass and age. According to the evolutionary tracks in Fig.\,\ref{hr_diag}, the A0--A3 spectral types will correspond to different stellar masses if stars are close to the ZAMS or near the TAMS.  

This trend does not seem to explain the bimodality observed in Figs\,\ref{distrib2} and \ref{over_rot}. According to Table\,\ref{per_mode}, the bimodality reaches its maximum (in terms of percentage of the full distribution) in the mass bin $2.55<M/\ensuremath{M_\odot} <3.05$, for which no obvious trend is seen in Fig.\,\ref{age-pec}.d.

\subsection{Evolution of the distributions of rotational velocities}
\label{evol_distr_vel} 
 \begin{figure*}[!htp]
\sidecaption
  \includegraphics[width=12cm,viewport=17 11 437 323,clip]{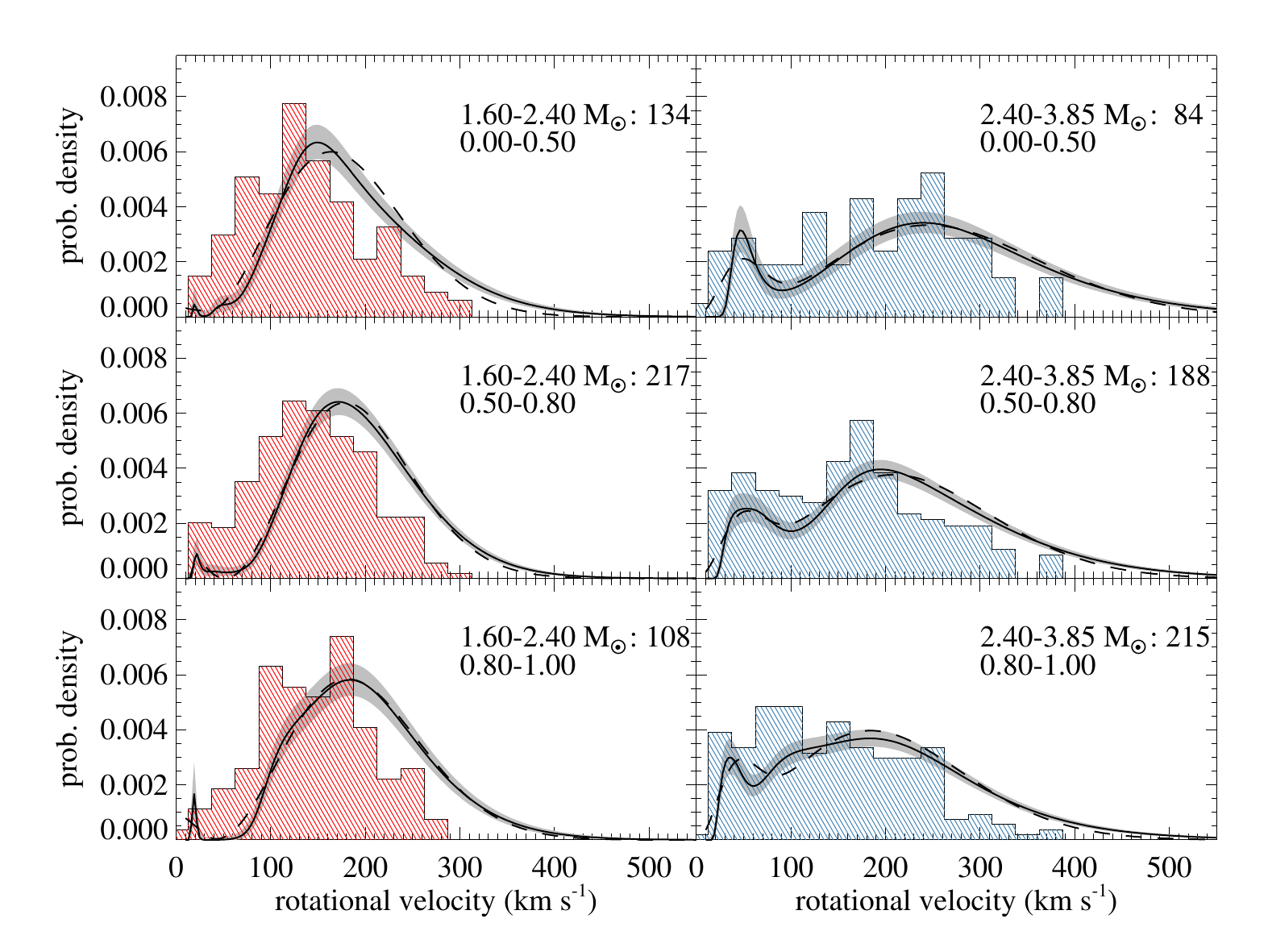}
     \caption{Distributions of rotational velocities as a function of age for two mass subsamples. Left panels correspond to the low-mass bin: 1.6--2.4\,\ensuremath{M_\odot}\ and right panels correspond to the high-mass bin: 2.4--3.85\,\ensuremath{M_\odot}. As in Fig.\,\ref{distrib2}, shaded histograms are the observed \ensuremath{v\sin i}; solid thick lines are the distributions of equatorial velocities \ensuremath{v}\ and the gray strips are their associated variability bands. The mass range for each subsample is indicated in the corresponding panel together with the number of stars, and below, the age interval is indicated and goes from young stars ($0<\ensuremath{t/t_\mathrm{MS}}<0.5$) in upper panels to stars close to the TAMS ($0.8<\ensuremath{t/t_\mathrm{MS}}<1$) in lower panels. The histograms are normalized to fit the probability density scale.}
     \label{distrib3}
\end{figure*}

  In the distributions shown in Fig.\,\ref{distrib2} stars are not distinguished by their ages. Below, stellar ages are taken into account in two different ways: (i) by studying the changes of the rotation velocity distributions as a function of time (this section); (ii) by studying the evolution from the ZAMS to the TAMS of the average rotational velocities per mass (Sect.~\ref{evol_rot_vel}).\par
  The evolution of the distributions of rotational velocity during the MS life time is studied separately for the two groups of stars identified in Fig.~\ref{over_rot}. The cut in mass was chosen to be $M\gtrsim2.4M_{\odot}$, however, to gather enough stars in the high-mass part.
 We obtained distributions of \ensuremath{v}\ for each mass group by separating the stars into three age intervals (Fig.\,\ref{distrib3}). They were chosen so that the number of stars in each of them enabled us to rectify the distributions from the projection angle effect. As seen in Fig.\,\ref{mass_age}, the young part of the diagram is poorly sampled. Hence the first bin is large, $0 <\ensuremath{t/t_\mathrm{MS}} <0.5$, to warrant statistical reliability. The two mass bins behave quite differently as a function of age. Table\,\ref{per_mode_age} gives the parameters that characterize the subsamples and their velocity distributions.

The low-mass star \ensuremath{v}\ distribution does not vary significantly in shape with age, and remains unimodal from ZAMS to TAMS. It shows an acceleration, however. The mode of the distribution increases from about 150\,\ensuremath{\mbox{km}\,\mbox{s}^{-1}}\ for $0<\ensuremath{t/t_\mathrm{MS}}<0.5$, to 170\,\ensuremath{\mbox{km}\,\mbox{s}^{-1}}\ for $0.5<\ensuremath{t/t_\mathrm{MS}}<0.8$ and to about 185\,\ensuremath{\mbox{km}\,\mbox{s}^{-1}}\ for $0.8<\ensuremath{t/t_\mathrm{MS}}<1$.

The high mass \ensuremath{v}\ distribution does exhibit significant changes as a function of age. The bimodality is very clear among stars younger than $\ensuremath{t/t_\mathrm{MS}}<0.8$, whereas the two peaks blend near the TAMS ($0.8<\ensuremath{t/t_\mathrm{MS}}<1$). Slowly rotating stars are present at any age, but the fast rotating part of the distribution decelerates with increasing age. In terms of positions of the mode, the distribution peaks at about 240\,\ensuremath{\mbox{km}\,\mbox{s}^{-1}}\ for the younger half ($\ensuremath{t/t_\mathrm{MS}}<0.5$), and stagnates at about 190\,\ensuremath{\mbox{km}\,\mbox{s}^{-1}}\ for the older part. In terms of median of the full distribution, however, the deceleration is continuous and the median $v$ decreases from 250\,\ensuremath{\mbox{km}\,\mbox{s}^{-1}}\ for $0<\ensuremath{t/t_\mathrm{MS}}<0.5$, to 210\,\ensuremath{\mbox{km}\,\mbox{s}^{-1}}\ for $0.5<\ensuremath{t/t_\mathrm{MS}}<0.8$ and to 185\,\ensuremath{\mbox{km}\,\mbox{s}^{-1}}\ for $0.8<\ensuremath{t/t_\mathrm{MS}}<1$.

 Using these distributions, the ratio of the average velocity of the fast rotating component (derived from the formula of the Maxwellian distribution) over the average velocity of the full component was derived. This ratio is characterized by the factor $\kappa_{\rm r}$, which allowed us to correct in Sect.~\ref{obs_vel_evol} the mean values of rotational velocities in mixed samples encompassing slowly and rapidly rotating stars. It corrects the curves of rotational velocities against stellar ages for the effect carried by the slowly rotating stars, so as to derive distributions and related modes for the rapidly rotating stars only.

 \begin{table*}[!htp]
\caption{Parameters of the 1D velocity distributions determined in the mass bins and age bins shown in Fig\,\ref{distrib3}.}
\centering
\begin{tabular}{cccc|cc|ccccccc}
\hline\hline
Mass & \ensuremath{t/t_\mathrm{MS}} & \#   &     & \multicolumn{2}{c|}{Slow rotators} & \multicolumn{6}{c}{Fast rotators} \\
range& range & stars& $\hat{h}$ &  $\mu_\mathrm{s}$  &  dispersion & \% & $\sqrt{2}\,\alpha$ & $\ell$ & $\mu_\mathrm{r}$ &$\mu^\prime_\mathrm{r}$& dispersion & $\kappa_\mathrm{r}$\\
(\ensuremath{M_\odot}) & & & & (\ensuremath{\mbox{km}\,\mbox{s}^{-1}})& (\ensuremath{\mbox{km}\,\mbox{s}^{-1}})& &(\ensuremath{\mbox{km}\,\mbox{s}^{-1}})&(\ensuremath{\mbox{km}\,\mbox{s}^{-1}}) & (\ensuremath{\mbox{km}\,\mbox{s}^{-1}})& (\ensuremath{\mbox{km}\,\mbox{s}^{-1}})& (\ensuremath{\mbox{km}\,\mbox{s}^{-1}})& \\
\hline
1.60--2.40&0.0--0.5&134&0.228&  --- & --- 	 &100 & 138$\pm 1$ &  26$\pm 1$ & 164&148& 65& 1.005\\   
1.60--2.40&0.5--0.8&217&0.204&  --- & --- 	 &100 & 127$\pm 1$ &  52$\pm 1$ & 179&172& 61& 0.979\\   
1.60--2.40&0.8--1  &108&0.236&  --- & --- 	 &100 & 141$\pm 1$ &  40$\pm 1$ & 181&184& 67& 0.995\\   
2.40--3.85&0.0--0.5& 84&0.317&  47 $\pm 1$ & 22  & 89 & 221$\pm 3$ &  24$\pm 3$ & 245&241&105& 1.101\\
2.40--3.85&0.5--0.8&188&0.268&  49 $\pm 1$ & 23  & 88 & 193$\pm 2$ &  15$\pm 3$ & 208&196& 92& 1.103\\
2.40--3.85&0.8--1  &215&0.276&  43 $\pm 1$ & 20  & 88 & 184$\pm 1$ &   0$\pm 1$ & 184&184& 88& 1.053\\
\hline
\end{tabular}
\label{per_mode_age}
\tablefoot{As in Table\,\protect\ref{per_mode}, the corresponding number of normal stars and the  smoothing parameter $\hat{h}$ (expressed in logarithmic \ensuremath{v\sin i}) are given for each subsample. For each distribution, the parameters of the Maxwellian fit are listed: percentage of fast velocity distributions, mode and dispersion. The mode of the slow rotator distribution $\mu_\mathrm{s}$ is derived from the Maxwellian fit. The fast rotator distribution is a lagged Maxwellian (see text and footnote\,\ref{lagmaxwell}) and its mode is $\mu_\mathrm{r}\equiv\sqrt{2}\alpha+\ell$. The estimator $\mu_\mathrm{r}^\prime$ was derived as the position of the maximum directly on the 1D fast distributions. Correction factors $\kappa_{\rm r}$ were derived as $\kappa_{\rm r}\equiv(2\,\alpha\sqrt{2/\pi}+\ell)/\langle v\rangle$, where $\langle v\rangle$ is the mean of the full distribution.}
\end{table*}

\section{Evolution of the rotational velocities compared with models}
\label{evol_rot_vel}

    In a first attempt, we tried to apprehend possible dependencies of the evolution with age of rotational velocities for different masses through the \ensuremath{v/v_\mathrm{crit}}\ velocity ratio, where \ensuremath{v_\mathrm{crit}}\ is the time- and mass-dependent critical equatorial rotational velocity. In a second approach, we studied the \ensuremath{v/v_\mathrm{ZAMS}}\ velocity ratio, where \ensuremath{v_\mathrm{ZAMS}}\ is the equatorial rotational velocity on the ZAMS.

    The velocity ratios \ensuremath{v/v_\mathrm{crit}}\ and  \ensuremath{v/v_\mathrm{ZAMS}}\ were compared with model predictions for three different regimes of internal redistribution of the angular momentum. The first two predictions are for extreme redistributions and are detailed in the present discussion. The third type of models were calculated by \citet{Ekm_08} for stars with $3\,\ensuremath{M_\odot}$, the only common mass with the present paper.

   The ZAMS was considered as the initial state of the MS stellar evolutionary phase. Nevertheless, as a consequence of all possible physical conditions prevailing during the evolutionary stages preceding the ZAMS, a given star on the ZAMS could be either a rigid rotator or a differential rotator. If the star is a differential rotator, the total angular momentum stored in the object can be higher or lower than the limiting value supported by a critical rigid rotator. The characteristics of the stellar structure depend on the amount of rotational energy stored and on the internal distribution of the angular velocity. The competition between gravitational and centrifugal forces determines the final stellar internal structure, the meridional circulation, and the mixing phenomena of chemical elements that put constraints to the subsequent stellar evolution in the MS \citep[cf.][]{MarMet00}.

   The mass-loss rates inferred for A-type stars are on the order of $10^{-12}$ to $10^{-10}$\,\ensuremath{M_\odot}\,yr$^{-1}$ \citep{LazCaa92,LasCai99}, which for evolutionary time scales ranging from $10^8$ to some $10^9$\, yr implies that the amount of angular momentum removed in these objects during their MS life span is probably negligible on average. We assumed then that all studied stars evolve during the MS phase conserving their total angular momentum.

     The amount of rotational kinetic energy stored by a star is currently estimated using the ratio $\tau=K/|W|$, where $K$ is the kinetic energy and $W$ is the gravitational potential energy. At rotational energies approaching $\tau\approx0.27$, the object is dynamically unstable. For $\tau\approx0.14$ it becomes secularly unstable, in the sense that its axial symmetry can be broken and transformed into a three-axially symmetric body \citep{Tal78}. If at any time during the MS evolution a star has a total angular momentum higher than permitted by the critical rigid rotation, the star is said to evolve as a \emph{neat} differential rotator. We refer to the evolution at low rotational energy regime if stars have $\tau\lesssim\tau^\mathrm{rigid}_\mathrm{crit}$, otherwise they will be said to evolve at high rotational energy. The notation $\tau^\mathrm{rigid}_\mathrm{crit}$ stands for the critical, or upper possible energy born by a rigid rotator, which ranges from $\tau\lesssim\tau^\mathrm{rigid}_\mathrm{crit}=0.0041$ to 0.0053 for stars with masses from $M=1.5$ to 3\,\ensuremath{M_\odot}, respectively. Assuming that stars are permanently axisymmetric, the evolution at high-energy regime implies that $\tau^\mathrm{rigid}_\mathrm{crit}\lesssim\tau\lesssim0.14$.

    In the present discussion we assumed that stars evolve in the MS at low rotational energy regime, i.e. they have $\tau\lesssim\tau^\mathrm{rigid}_\mathrm{crit}$. Moreover, in our calculations we still limited the total angular momentum to $J\leq J^\mathrm{TAMS}_\mathrm{crit}$, where for a given mass, $J^\mathrm{TAMS}_\mathrm{crit}$ corresponds to the angular momentum supported by a critical rigid rotator on the TAMS. This choice is mandatory for stars that are assumed to be rigid rotators on the ZAMS and they evolve in the MS by conserving the total angular momentum. In fact, a star steadily becomes more centrally condensed and its radius is enlarged as it evolves from the ZAMS to the TAMS. This decreases the upper limit of total angular momentum with time that the star can bear.

\begin{figure}[!htp]
\centering
\resizebox{\hsize}{!}{\includegraphics{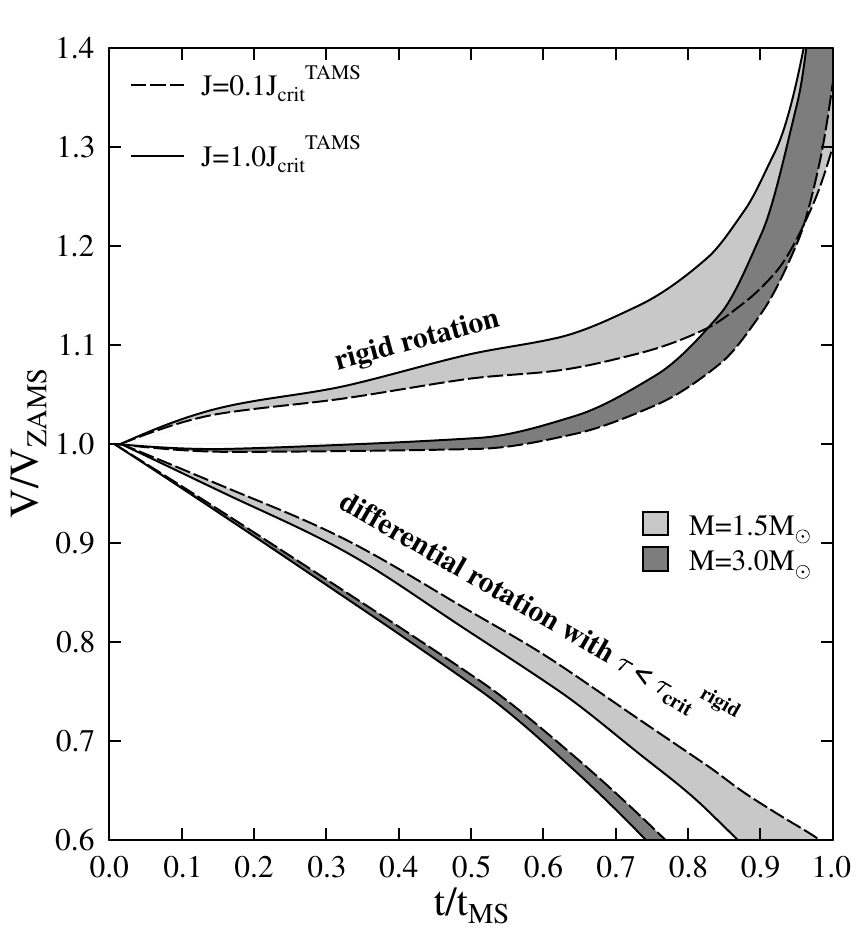}}
\caption{\ensuremath{v/v_\mathrm{ZAMS}}\ ratio of equatorial rotational velocities as a function of the \ensuremath{t/t_\mathrm{MS}}\ time ratio in the MS for stars with masses from $M=1.5$ to 3\,\ensuremath{M_\odot}. \ensuremath{v_\mathrm{ZAMS}}\ is the equatorial rotational velocity on the ZAMS, $t$ is the stellar age and $t_{\rm MS}$ is the time that a star of given mass $M$ can remain in the MS. Two extreme cases are represented: stars evolving as rigid rotators (instantaneous complete angular momentum redistribution) and as differential rotators without any angular momentum redistribution. Dashed lines represent the limits for $J/M\to0$, while the solid lines are for stars that on the ZAMS have a specific angular momentum that in the TAMS becomes the critical one.} 
\label{om_rig_diff}
\end{figure}

\begin{figure}[!htp]
\centering
\resizebox{\hsize}{!}{\includegraphics{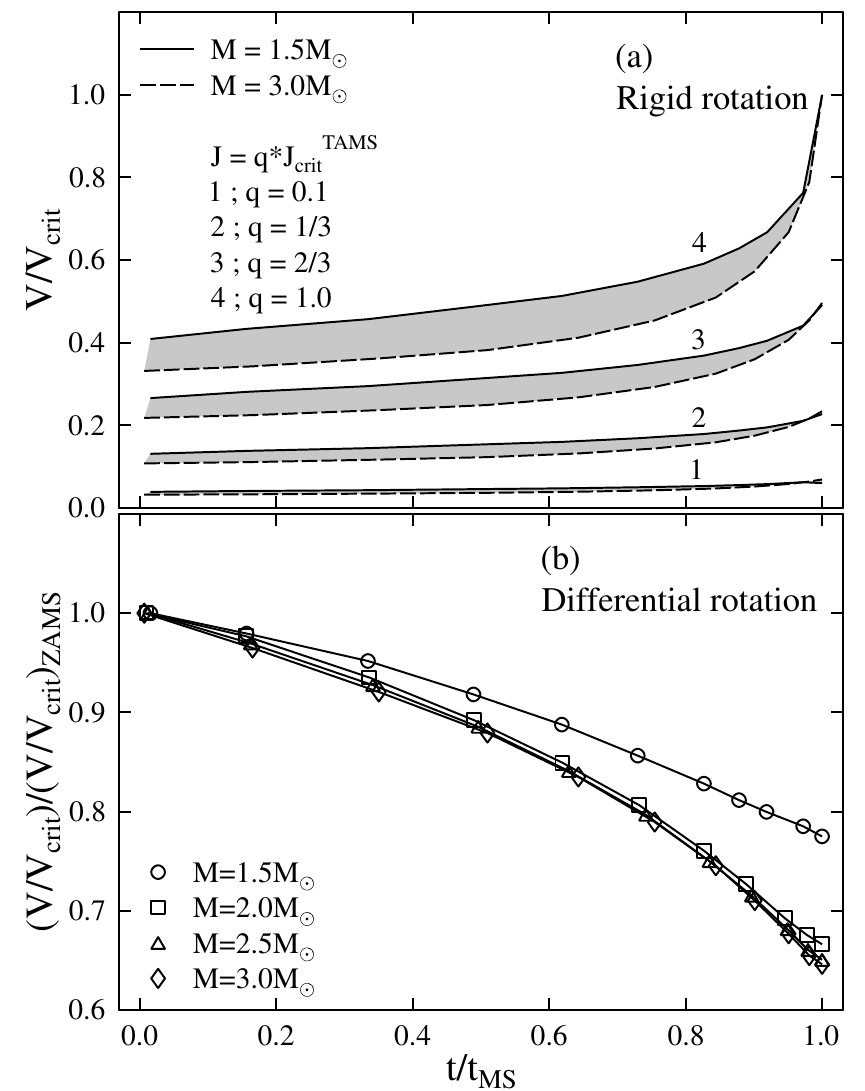}}
\caption{Theoretical evolution of the \ensuremath{v/v_\mathrm{crit}}\ velocity ratio calculated for several stellar masses. (a) Dependence with time of the \ensuremath{v/v_\mathrm{crit}}\ ratio for several values of the total specific angular momentum $J$, when stars evolve as rigid rotators all long the MS phase; (b) evolution of the ratio \ensuremath{v/v_\mathrm{crit}}\ normalized to its value on the ZAMS when rotators conserve the angular momentum by shells.}
\label{vevc}       
\end{figure}

\subsection{Two limiting cases for the evolution of the equatorial rotational velocities}
\label{limit_vel}

 Two simple and extreme cases of the evolution of the surface angular velocity can be considered, depending on the way the star redistributes its angular momentum during the MS evolutionary phase. For the limited values of rotational energy stored by the stars on the ZAMS stated above, $J\leq J^\mathrm{TAMS}_\mathrm{crit}$, this can happen in two ways: (i) by entirely redistributing its angular momentum at any instant; (ii) by impeding any exchange of angular momentum among the stellar shells, so that each of them conserves its specific angular momentum. In the first case the star evolves as a rigid rotator, while in the second case, we say the star evolves as a \emph{simple} differential rotator.

\subsubsection{Stars evolving as rigid rotators}
 The equatorial velocity at a time $t$ elapsed from the ZAMS ($t=0$) in a star 
rotating as a rigid rotator is roughly given by
\begin{equation}
\frac{\ensuremath{v}(t)}{\ensuremath{v_\mathrm{ZAMS}}} =\left[\frac{R(t)}{R_{\rm ZAMS}}\right]\left[\frac{I_{\rm ZAMS}}{I(t)}\right],
\label{veq_rig}
\end{equation}
where $[\ensuremath{v}(t),\ensuremath{v_\mathrm{ZAMS}}]$, $[R(t),R_{\rm ZAMS}]$ and $[I(t),I_{\rm ZAMS}]$ are the stellar equatorial linear rotational velocity, the radius and the moment of inertia at time $t$ and at the ZAMS, respectively. We calculated two-dimensional models of stars with rigid rotation as detailed in Appendix~\ref{mod_rig}. Figure\,\ref{om_rig_diff} shows the evolution of the ratio \ensuremath{v/v_\mathrm{ZAMS}}\ for stars with masses $M=1.5$ and 3\,\ensuremath{M_\odot}\ that conserve their total angular momentum during the MS evolutionary phase. The calculation was made for values of the angular momentum parametrized as $J=q\times J^\mathrm{TAMS}_\mathrm{crit}$, where $q$ takes the values 0.1 and 1. For objects undergoing total redistribution of the angular momentum, the velocity ratio \ensuremath{v/v_\mathrm{ZAMS}}\ increases very slowly or remains nearly constant over the first two-thirds of the MS evolutionary period. Only in the last third of the MS phase do the changes of the stellar structure as a function of the angular momentum produce some effects on the equatorial velocity, which slightly depend on the stellar mass.

  The variation of the \ensuremath{v/v_\mathrm{crit}}\ ratio in rigid rotators of different masses and with total angular momenta $J=q\times J^\mathrm{TAMS}_\mathrm{crit}$ ($q=0.1$, 1/3, 2/3, 1) is plotted in Fig.~\ref{vevc}a. The \ensuremath{v/v_\mathrm{crit}}\ velocity ratio of rigid rotators remains nearly unchanged during the first two-thirds of the MS evolutionary phase, and increases sensitively only in the last third of the MS. However, $(\ensuremath{v/v_\mathrm{crit}})_{\rm ZAMS}$ clearly depends on the total angular momentum $J$.

\subsubsection{Stars evolving as simple differential rotators}
\label{simplediffrot}
 The geometrical deformation of the stellar surface mainly depends on the specific angular momentum on the surface \citep{Zoc_11a,Zoc_11b}. Owing to the particular differential rotation assumed here, the specific angular momentum will become increasingly centrally condensed, so that the amount left in the surface will decrease strongly as the star evolves from ZAMS to TAMS. A simple account of the stellar deformation induced by the differential rotation could in principle be performed using a homoeoidal description of the internal density distribution \citep{Chr69}, where the specific angular momentum is constant over spheroidal shells. However, owing to the fairly small effects carried by the rotation studied here on the stellar shape, we assumed that stars are spherical.
 The angular momentum of a spherical shell of width $\mbox{d}r$ is
\begin{equation}
\mbox{d}J(t) = \left(\frac{8\pi}{3}\right)\rho(r)r^4\Omega(r)\mbox{d}r,
\label{veq_difj}
\end{equation}
where $\rho(r)$ and $\Omega(r)$ are the density and the angular velocity at radius $r$. $\Omega(r)$ is assumed to be uniform over each shell, i.e. ``shellular" distribution of the angular velocity. If $r$ and $r'$ are the radii at times $t$ and $t'$ of the same shell, whose mass $\mbox{d}M(r)=4\pi\rho(r)r^2\mbox{d}r$ and angular momentum are conserved, we have
\begin{equation}
\frac{\Omega'(r')}{\Omega(r)} = \left(\frac{r}{r'}\right)^2,
\label{veq_difom}
\end{equation}
which for the equatorial velocity at $t$ and $t'$ in the stellar surface implies that
\begin{equation}
\frac{\ensuremath{v}(t')}{\ensuremath{v}(t)} = \frac{R_{\rm e}(t)}{R_{\rm e}(t')},
\label{veq_difveq}
\end{equation}
where $R_{\rm e}(t)$ and $R_{\rm e}(t')$ are the equatorial radii of the star at times $t$ and $t'$, respectively. The change of the internal angular velocity as described by Eq.\,\ref{veq_difom} induces a small variation on the moment of inertia that is taken into account as described in Appendix~\ref{mod_diff}. The variation of the equatorial velocity with time thus obtained and normalized to its value on the ZAMS is plotted in Fig.~\ref{om_rig_diff}. The time dependent \ensuremath{v/v_\mathrm{ZAMS}}\ curves are also function of the stellar mass, while they are quite independent of the value of $J$.
\begin{figure}[!htp]
\centering
\resizebox{\hsize}{!}{\includegraphics{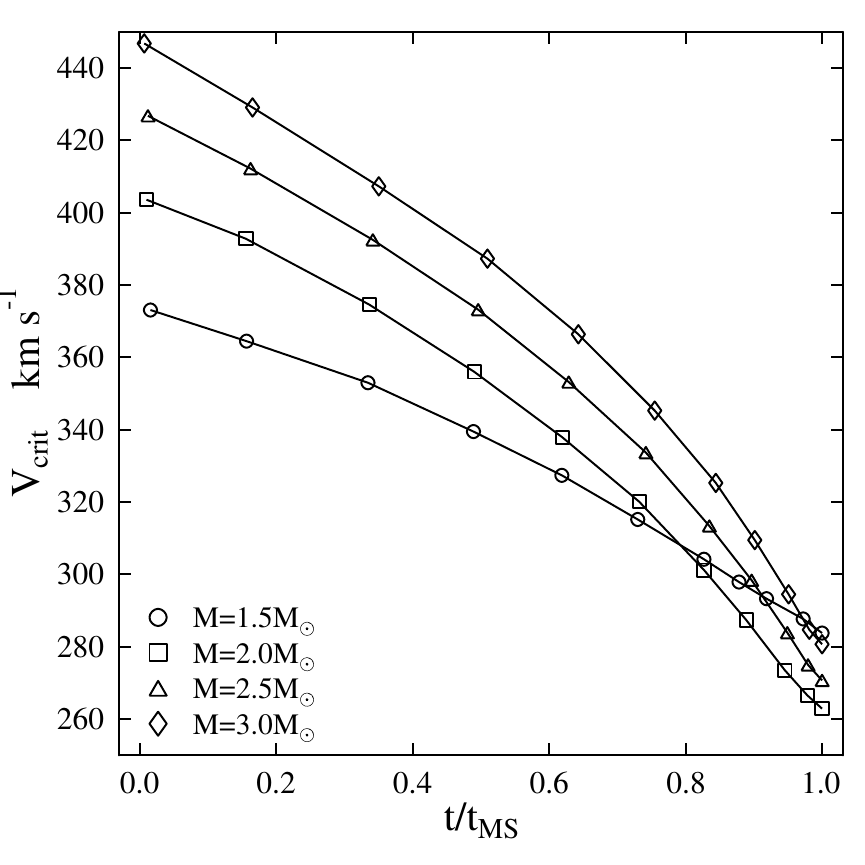}}
\caption{Critical equatorial rotational velocities of stars with different masses as a function of age in the MS. At each $t/t_{\rm MS}$ the objects are assumed to be rigid rotators.}
\label{crit_vel_fi}       
\end{figure}
\begin{figure*}[!htp]
\centering
\resizebox{\hsize}{!}{\includegraphics{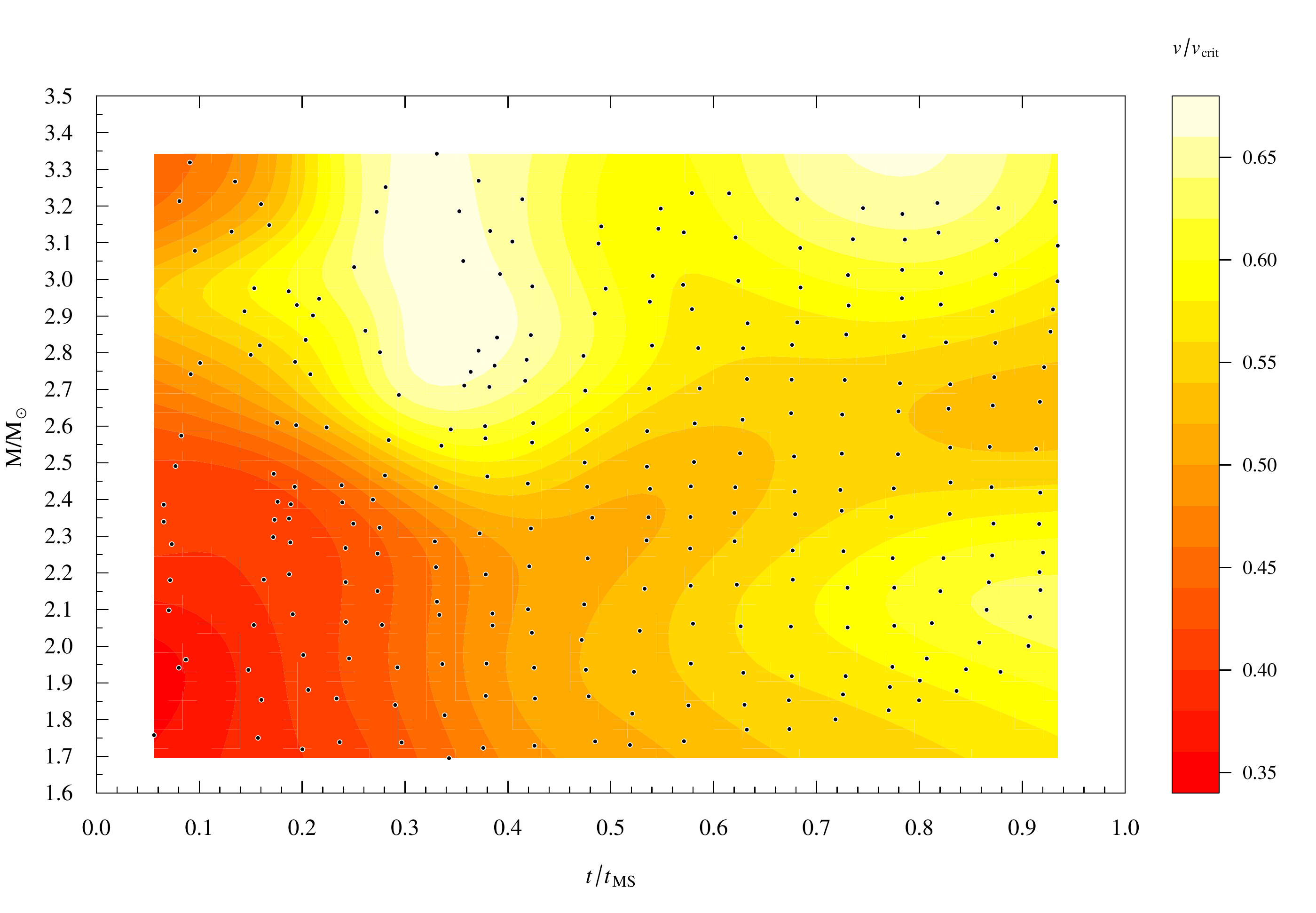}}
\caption{Distribution of the ratio \ensuremath{v/v_\mathrm{crit}}\ of true rotational velocities as a function of the fractional age in the MS evolutionary phase for normal stars in the mass interval $1.7\leq M\leq3.3\,\ensuremath{M_\odot}$. The filled circles are the average positions of the stars in running boxes with bins of 0.15 in \ensuremath{t/t_\mathrm{MS}}\ and 0.5 in $M/\ensuremath{M_\odot}$.}       
\label{distrib-age}
\end{figure*}

  In Fig.~\ref{vevc} we show the variation of \ensuremath{v/v_\mathrm{crit}}\ as a function of time for stars with masses from $M=1.5$ to 3\,\ensuremath{M_\odot}, and several values of the angular momentum $J=q\times J^\mathrm{TAMS}_\mathrm{crit}$ with $q$ ranging from 0.1 to 1. Although the ratio \ensuremath{v/v_\mathrm{crit}}\ depends on the value of $J$, its variation with time has a tiny dependence with mass. To demonstrate this small dependence with mass, the curves in Fig.~\ref{vevc}b are normalized to the respective 
values on the ZAMS: $[v(M)/v(M)_{\rm crit}]_{\rm ZAMS}$. Owing to the low values of $J$ chosen here, model stars can never have a critical specific angular momentum in the equator. For internal rotation laws given by Eq.\,\ref{veq_difom} this can happen only for much higher values of $J$, which means that in this case the stars behave in the MS always as \emph{neat} differential rotators.

\subsection{Observed evolution of rotational velocities vs. predicted extreme angular momentum
redistribution regimes}
\label{obs_vel_evol}

\begin{figure}[!htp]
\centering
\resizebox{\hsize}{!}{\includegraphics{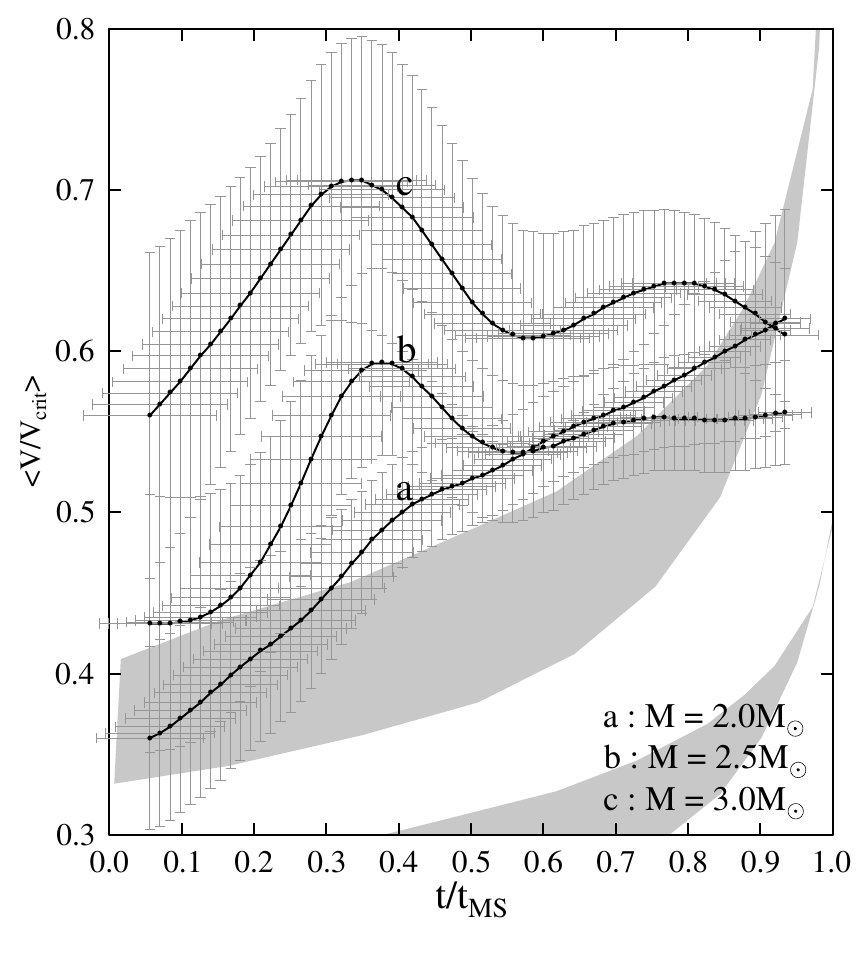}}
\caption{Evolution of the equatorial velocity ratio $\langle\ensuremath{v/v_\mathrm{crit}}\rangle$ in the MS life time span. The ratios of true equatorial velocities are calculated for three masses: $2\,\ensuremath{M_\odot}$, $2.5\,\ensuremath{M_\odot}$ and $3\,\ensuremath{M_\odot}$. The shaded regions correspond to the theoretical evolution of the \ensuremath{v/v_\mathrm{crit}}\ ratio in rigid rotators plotted in Fig.~\ref{vevc}, whose total angular momenta are $J=J^\mathrm{TAMS}_\mathrm{crit}$ (middle) and $J=(2/3)J^\mathrm{TAMS}_\mathrm{crit}$ (lower right corner).} 
\label{vel_crit}      
\end{figure}

\begin{figure}[!htp]
\centering
\resizebox{\hsize}{!}{\includegraphics{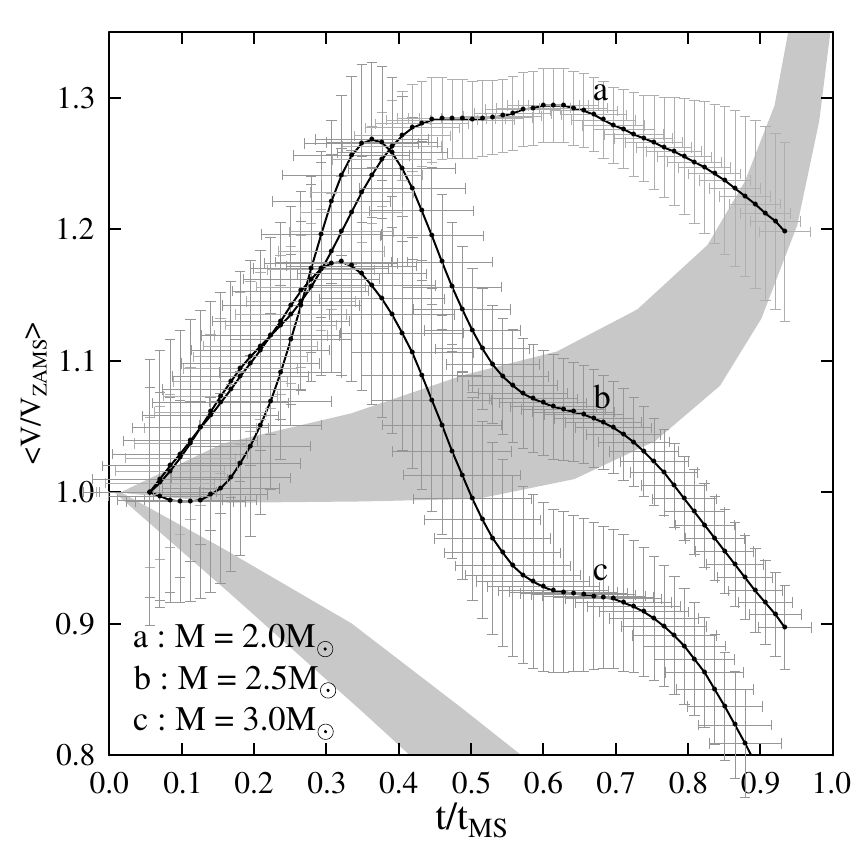}}
\caption{Evolution of the equatorial velocity ratio $\langle\ensuremath{v/v_\mathrm{ZAMS}}\rangle$ in the MS life time span, where \ensuremath{v_\mathrm{ZAMS}}\ is the true equatorial velocity of stars on the ZAMS. The average true equatorial velocities were calculated for the same masses as in Fig.~\ref{vel_crit}. The shaded regions correspond to the theoretical evolution of the \ensuremath{v/v_\mathrm{ZAMS}}\ ratios shown in Fig.~\ref{om_rig_diff}: rigid rotators (middle), simple differential rotators (lower left corner).}
\label{om_rig_dif_obs}       
\end{figure}

  A general picture derived using observations of the evolution of rotational velocities in the MS as a function of mass is given in Fig.~\ref{distrib-age}, where the average ratio of true rotational velocities $\langle\ensuremath{v/v_\mathrm{crit}}\rangle$  is shown as a function of the fractional age $t/t_{\rm MS}$ in the MS for masses in the interval $1.7\leq M\leq3.3\,\ensuremath{M_\odot}$. \ensuremath{v_\mathrm{crit}}\ is the equatorial critical velocity given by
\begin{equation}
\ensuremath{v_\mathrm{crit}} = 436.7\left(\frac{M}{\ensuremath{M_\odot}}\right)^{1/2}\left[\frac{R_{\odot}}{R_{\rm crit}(M,t)}\right]^{1/2} \quad \ensuremath{\mbox{km}\,\mbox{s}^{-1}},
\label{crit_vel_fo}
\end{equation} 
where $R_{\rm crit}(M,t)$ is the stellar radius at critical rotation. For each star, the mass- and time-dependent critical radius $R_{\rm crit}(M,t)$ was determined using two-dimensional models of rigidly rotating stars, whose characteristics are given in Appendix~\ref{mod_rig}. Figure~\ref{crit_vel_fi} plots $\ensuremath{v_\mathrm{crit}}(M,t)$ against $t/t_{\rm MS}$ for several stellar masses ranging from $M=1.5$ to $3\,\ensuremath{M_\odot}$.

  To obtain the diagram of average true rotational velocities $\langle\ensuremath{v/v_\mathrm{crit}}\rangle$ shown in Fig.~\ref{distrib-age}, each observed \ensuremath{v\sin i}\ parameter was divided by the corresponding $\ensuremath{v_\mathrm{crit}}(M,t)$. The values were averaged in running boxes with age bins of 0.18 for $\ensuremath{t/t_\mathrm{MS}}<0.3$ and 0.15 for $\ensuremath{t/t_\mathrm{MS}}>0.3$ and mass bins of 0.5\,\ensuremath{M_\odot}. These intervals are wide enough to ensure that the number of stars entering each average leads to statistically reliable transformations from $\langle\ensuremath{v\sin i}\rangle$ to $\langle\ensuremath{v}\rangle$ and they are narrow enough to warrant that useful information on the evolutionary characteristics of rotation is hardly not obliterated. The true equatorial velocities come from the average of the \ensuremath{v\sin i}\ parameter corrected for $\sin i$ according to the statistical principles given by \citet{ChrMuh50}. The two-dimensional distribution was smoothed with a smoothing parameter derived as in Sect.\,\ref{dist_vel} \citep[see][]{BonAzi97}.

  The two mass regions already identified in Fig.\,\ref{over_rot} depict in Fig.~\ref{distrib-age} an evolution of $\langle\ensuremath{v/v_\mathrm{crit}}\rangle$ with different characteristics. The limit between these regions is roughly at $M\approx2.5\,\ensuremath{M_\odot}$. To obtain a better insight on the characteristics of the evolution of the $\langle\ensuremath{v/v_\mathrm{crit}}\rangle$ ratio in each of these mass-intervals, constant mass curves were extracted from Fig.~\ref{distrib-age} for three stellar masses: 2, 2.5 and $3\,\ensuremath{M_\odot}$, which represent the average evolution of the ratio $\langle\ensuremath{v/v_\mathrm{crit}}\rangle$ as a function of time $t/t_{\rm MS}$ and also describe the most typical aspects observed of the evolution of $\langle\ensuremath{v/v_\mathrm{crit}}\rangle$. These curves can be taken as tracers of the evolution of the average true rotational velocity against the fractional age $t/t_{\rm MS}$. However, for masses  $M\ge2.5\,\ensuremath{M_\odot}$, these curves carry information on the evolution of two mixed groups of stars that we have already differentiated into slowly and rapidly rotating populations, of which the last is the most numerous one. Because the evolution of the rotational velocity in these stellar groups can present systematic differences, we aimed at obtaining the $\langle\ensuremath{v/v_\mathrm{crit}}\rangle$ curves that represent only the rapidly rotating stellar population. 
To this end, we used the correction factor $\kappa_\mathrm{r}$ (Table\,\ref{per_mode_age}) that converts the mean values $\langle v_{\rm s+r}\rangle$ of mixed slowly and rapidly rotating stars ``s+r'' to the searched rapid one $\langle v_{\rm r}\rangle=\kappa_{\rm r}(t)\times\langle v_{\rm s+r}\rangle$. The values calculated for the mid-time of each age interval were then interpolated for the required ages in the time interval $0\leq t/t_{\rm MS}\leq1$. 
Because $\kappa_{\rm r}(t)$ remains fairly constant over the first two thirds of the MS lifetime span, and the curves of average rotational velocities that we study here were normalized to the average velocity on the ZAMS, only the $\langle v\rangle$ velocities in the last evolutionary phases on the MS do undergo some little changes as compared to the original mixed average velocity distributions.

 The corrected $\langle\ensuremath{v/v_\mathrm{crit}}\rangle$ curves as a function of the fractional age $t/t_{\rm MS}$ are presented in Fig.~\ref{vel_crit}, where the vertical error bars give an insight on the uncertainties affecting the $\langle\ensuremath{v/v_\mathrm{crit}}\rangle-$ratio determination and the horizontal error bars are the averaged $\sigma_{\ensuremath{t/t_\mathrm{MS}}}$ (from Table\,\ref{fund_par}) in each bin. In this figure are superimposed the bands that correspond to the evolution of \ensuremath{v/v_\mathrm{crit}}\ ratio for rigid rotators of masses from $M=1.5$ to $3\,\ensuremath{M_\odot}$ previously shown in Fig.~\ref{vevc}. They represent two angular momenta: $J=J^\mathrm{TAMS}_\mathrm{crit}$ (middle of the diagram) and $J=(2/3)J^\mathrm{TAMS}_\mathrm{crit}$ (lower right corner). 
Even if in Fig.~\ref{vel_crit} the evolution of simple differential rotators is not shown, we notice that the evolution of $\langle\ensuremath{v/v_\mathrm{crit}}\rangle$ differs strongly from what is predicted from the two extreme possibilities of angular momentum redistribution described in Sect.~\ref{limit_vel}. In fact, in the first third of the MS and for both mass-groups, the ratio $\langle\ensuremath{v/v_\mathrm{crit}}\rangle$ increases faster than suggested by the theoretical predictions. The $\langle\ensuremath{v/v_\mathrm{crit}}\rangle$ ratio of stars with $M=2\,\ensuremath{M_\odot}$ increases monotonically until the TAMS, while in stars with $M\ge2.5\,\ensuremath{M_\odot}$ it reaches a maximum at $t/t_{\rm MS}\approx 0.3$. There is then a decrease that lasts roughly $\Delta t/t_{\rm MS}\approx0.2$, followed by a uniform value of $\langle\ensuremath{v/v_\mathrm{crit}}\rangle$ until the TAMS. 

    We calculated the mass- and time-dependent \ensuremath{v_\mathrm{crit}}\ corresponding to all points in the curves of Fig.~\ref{vel_crit} and derive the evolution of the average true equatorial velocities $\langle\ensuremath{v}\rangle$ normalized to $\langle \ensuremath{v}\rangle_{\rm ZAMS}$ for the same masses. The result is plotted in Fig.~\ref{om_rig_dif_obs}, where we also show the respective uncertainties. In this figure are superimposed the sequences of model $v/v_{\rm ZAMS}$ variations for rigid (middle of the figure), and for simple differential rotators (lower left corner) previously shown in Fig.~\ref{om_rig_diff}.

  The common characteristic of $\langle\ensuremath{v/v_\mathrm{ZAMS}}\rangle$ curves seen in Fig.~\ref{om_rig_dif_obs} is that for all masses they increase fast in the first third of the MS phase. For stars with $M\ge2.5\,\ensuremath{M_\odot}$ (curves b and c), the ratio $\langle\ensuremath{v/v_\mathrm{ZAMS}}\rangle$ decreases more or less monotonically in the second half of the MS phase. Owing to the shown uncertainties, we cannot be sure that the slopes reveal a slightly faster decrease than predicted for simple differential rotators. If it were actually the case, this decrease could imply some redistribution of angular momentum toward the center of stars, provided that in the meantime no strong loss of angular momentum through stellar winds occurs. For stars with $M=2\,\ensuremath{M_\odot}$ (curve a), $\langle\ensuremath{v/v_\mathrm{ZAMS}}\rangle$ remains almost constant from $t/t_{\rm MS}\approx0.4$ until $t/t_{\rm MS}\approx0.7$, and decreases slowly as the evolution approaches the end of the MS phase. This decrease follows the slope of simple differential rotators very closely, nearly as if the stars were not undergoing any redistribution of the angular momentum in the external layers during the final stages of their MS phase.

\begin{figure}[!htp]
\centering
\resizebox{\hsize}{!}{\includegraphics{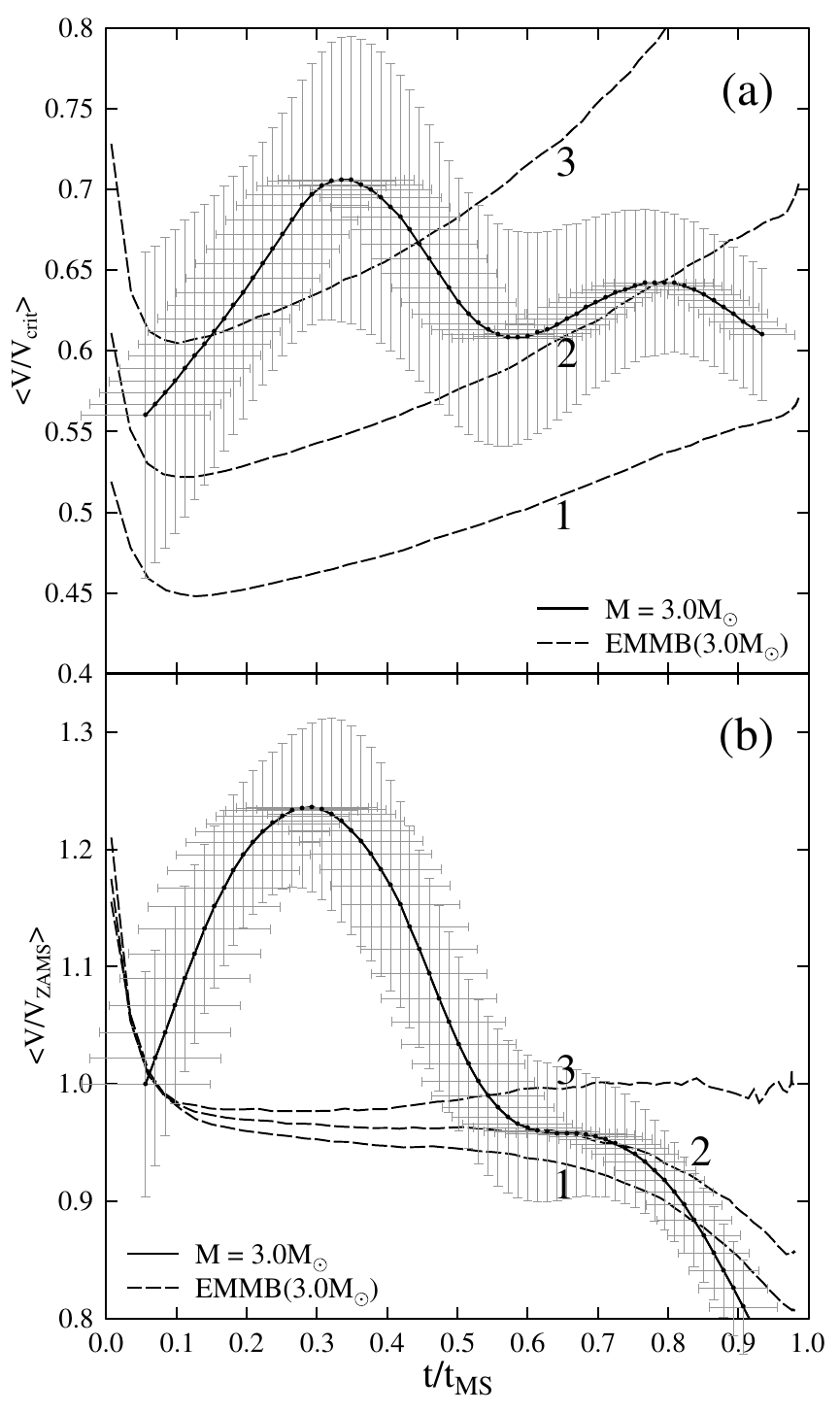}}
\caption{Comparison of the observed evolution of equatorial velocity ratios in the MS life time span for stars with $M=3\,\ensuremath{M_\odot}$ inferred from observations with theoretical ones calculated by \citet{Ekm_08} for different \ensuremath{v/v_\mathrm{crit}}\ on the ZAMS. (a) Evolution of $\langle\ensuremath{v/v_\mathrm{crit}}\rangle$ ratios; (b) evolution of $\langle\ensuremath{v/v_\mathrm{ZAMS}}\rangle$ ratios.}
\label{obs_emmb}       
\end{figure}

 In Figs.~\ref{vel_crit}, \ref{om_rig_dif_obs}, and \ref{obs_emmb} we show the uncertainties associated to the averages of fractional ages \ensuremath{t/t_\mathrm{MS}}, which in all cases are $\sigma_{\ensuremath{t/t_\mathrm{MS}}}\lesssim0.1$. In Table~\ref{mass} we see that only for $\ensuremath{t/t_\mathrm{MS}}\gtrsim0.5$, systematic deviations in the age estimations can be introduced by effects caused by the rapid rotation if $\Omega/\Omega_{\rm c}\gtrsim0.8$. However, because we note from Fig.~\ref{vel_crit} that $\langle \ensuremath{v/v_\mathrm{crit}}\rangle\pm\sigma_{\ensuremath{v/v_\mathrm{crit}}}\lesssim0.7$ in $\ensuremath{t/t_\mathrm{MS}}\gtrsim0.5$, they probably do not strongly affect the description of the evolution of rotational velocities obtained here.

\subsection{Observed evolution of rotational velocities vs. detailed calculations of the internal angular-momentum redistribution}
\label{obs_vel_evoldet}

\citet{Ekm_08} computed stellar models to derive the evolution of surface velocities during the MS lifetime. 
Figure~\ref{obs_emmb}a displays the comparison of the observationally inferred evolution of the \ensuremath{v/v_\mathrm{crit}}\ ratio with those obtained by \citet{Ekm_08} for three solar mass stars with different initial \ensuremath{v/v_\mathrm{crit}}\ ratios. The theoretical \ensuremath{v/v_\mathrm{crit}}\ ratios reveal an initial short period where a star redistributes its internal angular momentum passing from a rigidly rotating object to a differential rotator at the low-energy regime \citep{MarMet00}. We must note that the observed curve corresponds to a mixed stellar population where the average mass is $M=3\,\ensuremath{M_\odot}$. The averaging of velocities is necessarily made over a distribution of initial velocities that we do not know. We could assume some kind of Maxwellian or Gaussian initial distribution of equatorial velocities to obtain curves for single initial \ensuremath{v/v_\mathrm{crit}}\ ratios using perhaps some kind of deconvolution. The operation would nevertheless lead to uncertain results. However, the differences shown in Fig.~\ref{obs_emmb}a are eloquent enough by themselves to indicate that in the first half of the MS evolutionary phase the angular momentum in actual stars may undergo other redistribution processes than those theoretically predicted today. This difference appears still more clearly in Fig.~\ref{obs_emmb}b, where we compare the observed evolution of \ensuremath{v/v_\mathrm{ZAMS}}\ ratios with theoretical curves for the same initial velocity ratios as in Fig.~\ref{obs_emmb}a. In this figure the incidence of initial velocity distributions would have hardly any effect in the first half of the MS evolutionary phase, so that the above noted discrepancy between theory and observation seems to be confirmed.

  Excepting the very first drop of velocity ratios to be caused by an initial fast internal angular redistribution, we note that in the time interval $0.1\lesssim \ensuremath{t/t_\mathrm{MS}}\lesssim0.6$ the theoretical curves in Fig.~\ref{obs_emmb}b differ very little from the evolution of rigid rotators. In the last MS evolutionary span, the detailed calculations suggest a lowering of the equatorial velocity whose steepness is flatter than for simple differential rotators, which indicates that the stellar surface layer gains some angular momentum from that stored in the stellar interior.

  In all the above comparisons we note that for stars with masses $M\ge2.5\,\ensuremath{M_\odot}$ there seems to exist a characteristic step or time scale at which significant changes of $\langle\ensuremath{v}\rangle$ are produced that do not seem to be present in the theoretical models, which amounts to
\begin{equation}
\delta t \approx 0.2\,t_{\rm MS}.
\label{tscale}
\end{equation} 

\subsection{Discussion of the possible internal distribution of the angular velocity}
\label{int_om}

   A different insight in the evolution of rotational velocities can be obtained by comparing the minimum total specific angular momentum $J/M$ required to account for the equatorial velocities shown in Fig.~\ref{om_rig_dif_obs}. The minimum angular momentum $J/M$ is meant here to be the value stored in a rigid rotator that accounts for the observed rotational velocity $v$ at a given $t/t_{\rm MS}$. Let us then assume that the sequences of $\langle\ensuremath{v/v_\mathrm{ZAMS}}\rangle$ plotted in Fig.~\ref{om_rig_dif_obs} represent the evolution of the rotational velocity in actual single stars with the masses indicated in the figure. Because the mass loss rate in these stars is very low, we may consider that they evolve with conservation of the total angular momentum all along their MS phase. For all masses in Fig.~\ref{om_rig_dif_obs} and for the following stages $t/t_{\rm MS}\approx0.07$ (near the ZAMS), $t/t_{\rm MS}\approx0.3$--0.4 (when $v$ is maximum for stars with $M\gtrsim2.5\,\ensuremath{M_\odot}$), $t/t_{\rm MS}\approx0.65$ (near maxima and/or inflections of $v$) and $t/t_{\rm MS}\approx0.9$ (near the TAMS), we calculated the total specific angular momentum $J/M$ that rigid rotators require to account for the observed equatorial velocities. We also estimated the angular momentum of the critical rigid rotators at each chosen evolutionary stage. The estimated values of $J/M$ are given in Table~\ref{vel_synop} and the results show that
\begin{enumerate}[(i)]
\item for all evolutionary stages and in all studied masses we have on average $J/M<(J/M)^\mathrm{ZAMS}_\mathrm{crit}$, but roughly $J/M\gtrsim0.5\times(J/M)^\mathrm{ZAMS}_\mathrm{crit}$);
\item for almost all masses and evolutionary stages ranging from $t/t_{\rm MS}=0.3$ to 0.6, we notice that $J/M>(J/M)_{\rm ZAMS}$, which indicates that to account for the rotational velocity of stars at these stages, more rotational kinetic energy is needed than we inferred when we assume that they are mere rigid rotators on the ZAMS;
\item for masses $M=2\,\ensuremath{M_\odot}$ we find that $J/M<(J/M)^\mathrm{TAMS}_\mathrm{crit}$ in all evolutionary stages, while for higher masses stars need $J/M>(J/M)^\mathrm{TAMS}_\mathrm{crit}$ to account for their surface rotation at $t/t_{\rm MS}\approx0.3$.
\end{enumerate}

  We can then comment that
\begin{enumerate}[(i)]
\setcounter{enumi}{3}
\item the condition $J/M<(J/M)_\mathrm{crit}^\mathrm{ZAMS}$ in all evolutionary stages means that the evolution of rotational velocities in the MS of stars in the studied mass ranges can be consistent with low regimes of rotational energies, i.e. $\tau=K/|W|<\tau(${\tiny ZAMS})$_{\rm crit}^{\rm rigid}$;
\item knowing that for all evolutionary stages (including the ZAMS) the total specific angular momentum is estimated supposing that stars are rigid rotators, values $J(t)/M>(J/M)_{\rm ZAMS}$ at any $t/t_{\rm MS}>0$ imply that there must be some transfer of angular momentum from the center toward the surface. However, because rigid rotation in the initial evolutionary phases demands that $\Omega_{\rm core}=\Omega_{\rm envelope}$, in latter evolutionary phases where $J/M>(J/M)_{\rm ZAMS}$ would necessarily imply $\Omega_{\rm core}<\Omega_{\rm envelope}$, which could be not realistic. An increase of $\Omega_{\rm envelope}$ could then be possible if stars actually evolve as differential rotators having $\Omega_{\rm core}>\Omega_{\rm envelope}$ since the ZAMS. However, the condition $J/M<(J/M)^\mathrm{ZAMS}_\mathrm{crit}$ suggests that they could be \emph{simple} differential rotators on the ZAMS;
\item because in the studied range of stellar masses the total angular momentum of stars is conserved during the MS evolutionary phase, in the stages where it is inferred that $J/M>(J/M)^\mathrm{TAMS}_\mathrm{crit}$, stars have to redistribute their total angular momentum and end their MS phase behaving as neat differential rotators.
\end{enumerate}

\begin{table}[!htp]
\caption[]{Synopsis of the average evolution of equatorial velocities of $2\,\ensuremath{M_\odot}$, $2.5\,\ensuremath{M_\odot}$ and $3\,\ensuremath{M_\odot}$ stars.}
\centering
\tabcolsep 4.5pt
\begin{tabular}{ccc|cc}
\hline
\hline	
 $t/t_{\rm MS}$ & \ensuremath{v}\          &  $J/M$  & \ensuremath{v_\mathrm{crit}}\ &  $J/M$ \\
                & (\ensuremath{\mbox{km}\,\mbox{s}^{-1}})  &  ($10^{17}$cm$^2$\,s$^{-1}$)   &   (\ensuremath{\mbox{km}\,\mbox{s}^{-1}})     & ($10^{17}$cm$^2$\,s$^{-1}$)  \\
\hline	
\multicolumn{5}{c}{$M=2\,\ensuremath{M_\odot}$} \\
\hline 
0.07 & 147 & 0.782 & 404 & 1.592 \\
0.33 & 179 & 0.880 & 376 & 1.479 \\
0.62 & 188 & 0.891 & 346 & 1.305 \\
0.90 & 176 & 0.747 & 288 & 0.941 \\
\hline	
\multicolumn{5}{c}{$M=2.5\,\ensuremath{M_\odot}$} \\
\hline 
0.07 & 182 & 1.086 & 422 & 1.974 \\
0.36 & 231 & 1.270 & 391 & 1.842 \\
0.62 & 195 & 1.154 & 361 & 1.652 \\
0.90 & 167 & 0.875 & 299 & 1.201 \\
\hline	
\multicolumn{5}{c}{$M=3\,\ensuremath{M_\odot}$} \\
\hline 
0.07 & 249 & 1.558 & 439 & 2.389 \\
0.29 & 290 & 1.725 & 410 & 2.243 \\
0.62 & 229 & 1.419 & 376 & 2.017 \\
0.90 & 192 & 1.157 & 310 & 1.431 \\
\hline	
\end{tabular}
\label{vel_synop}
\end{table}

  Noting from the above arguments that stars can behave as differential rotators, we can try to speculate
\begin{enumerate}[(A)] 
\item what the internal rotation law of a star on the ZAMS can look like, if we assume that its total specific angular momentum is the highest we can find in a given MS evolutionary stage as shown in Table~\ref{vel_synop};\label{Q1}
\item what the required values of $J/M$ and the associated internal rotation laws can be to account for the observed average rotational velocity at the spotted evolutionary stages in Table~\ref{vel_synop}, if the star is assumed to be a differential rotator, and what the possible internal rotation laws on the ZAMS are that can afford the same $J/M$ values to explain the average ZAMS equatorial velocities.\label{Q2}
\end{enumerate}
  The ratio $\Omega_\mathrm{o}/\Omega_{\rm e}$ strongly depends on the characteristics of the internal rotational law ($\Omega_\mathrm{o}$ is the angular velocity of the center of the star, $r = R/R_{\rm e}=0$; $\Omega_{\rm e}$ is the angular velocity in the equator at the surface, $r = R/R_{\rm e}=1$). These notations should not be confused with $\Omega_{\rm core}$ and $\Omega_{\rm envelope}$, which vaguely refer to angular velocities in larger domains in the core and in the stellar envelope, respectively. Because at the moment we do not know what this law can be, we adopted the model-laws inferred by \citet{MarMet00} for early-type stars in the MS to make an educated guess on the possible ratios $\Omega_\mathrm{o}/\Omega_{\rm e}$. These laws can be sketched analytically as follows \citep{Zoc_07b}
\begin{equation}
\Omega(r) = \Omega_\mathrm{o}[1-p\times\exp(-a/r^b)],
\label{omega_r}
\end{equation}
where $\Omega(r)$ is uniform over each spherical shell of radius $r$, i.e. ``shellular" internal distribution of the angular velocity; from Eq.\,\ref{omega_r} it is $\Omega_{\rm e}=\Omega(r=1)$. The quantity $p$ is a contrast parameter; $a$ is a parameter that depends on the radius of the stellar core, which is assumed to be rotating rigidly; $b$ describes the steepness of the change of $\Omega(r)$ from $\Omega_\mathrm{o}$ to $\Omega_{\rm e}$. We note that $p=0$ is suited for rigid rotation and $p=1$ leads to the strongest possible $\Omega_\mathrm{o}/\Omega_{\rm e}$ ratio for Rayleigh's stability condition $\partial j/\partial r>0$ to be satisfied. The values of $b$ range from $b\approx3.5$ on the ZAMS to $b\gtrsim2$ in the TAMS \citep{MarMet00}. To make quantitative estimates, let us adopt a $2.5\,\ensuremath{M_\odot}$ test star with the rotational characteristics displayed in Table\,\ref{vel_synop}. The results obtained in the frame of the above quoted hypotheses (\ref{Q1}) and (\ref{Q2}) are given in Table~\ref{omc_ome}.

  According to question (\ref{Q1}), we see that for the $2.5\,\ensuremath{M_\odot}$ star the highest $J/M$ value in the MS occurs at $\ensuremath{t/t_\mathrm{MS}}=0.36$, where a rigid rotator needs $(J/M)_{\ensuremath{t/t_\mathrm{MS}}=0.36}=1.27\times10^{17}$\,cm$^2$\,s$^{-1}$, which is also the minimum value possible for $(J/M)$ to account for the observed average equatorial velocity $\ensuremath{v}=231$\,\ensuremath{\mbox{km}\,\mbox{s}^{-1}}\ at this evolutionary stage. Assuming that we have $(J/M)_{\rm ZAMS}=(J/M)_{\ensuremath{t/t_\mathrm{MS}}=0.36}$, using the rotation law in Eq.\,\ref{omega_r} we obtain the contrast parameter $p_{\rm ZAMS}=0.25$ and see
that $(\Omega_\mathrm{o}/\Omega_{\rm e})_{\rm ZAMS}=1.31$ to explain $\ensuremath{v}=182$\,\ensuremath{\mbox{km}\,\mbox{s}^{-1}}\ on the ZAMS.
 
\begin{figure}[!htp]
\centering
\resizebox{\hsize}{!}{\includegraphics{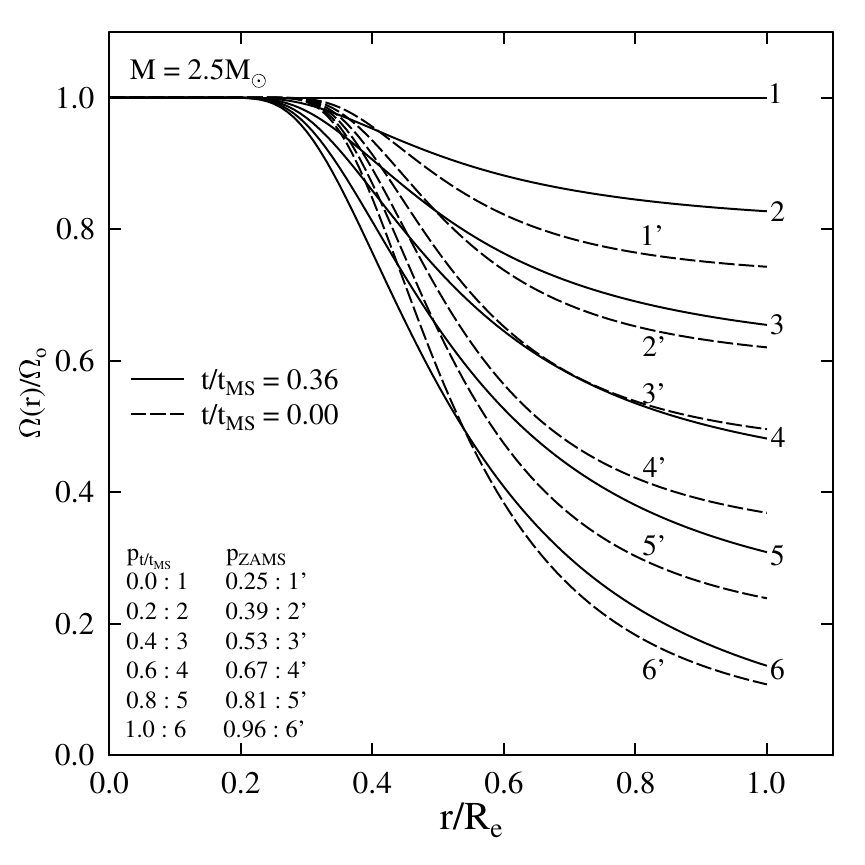}}
\caption{Internal rotation laws $\Omega(r)/\Omega_\mathrm{o}$ sketched with Eq.\,\ref{omega_r} and inferred for a $2.5\,\ensuremath{M_\odot}$ test star at $t/t_{\rm MS}=0.36$ and on the ZAMS, according to several parametrized values of the contrast parameter $p$ adopted for $t/t_{\rm MS}=0.36$ that all account for the observed rotational velocity at this evolutionary phase. For each parameter $p_{t/t_{\rm MS}}$ we obtained the corresponding $p_{\rm ZAMS}$, so that for the same specific total angular momentum $J/M$ the corresponding angular velocity ratio $\Omega_\mathrm{o}/\Omega_{\rm e}$ can explain the observed rotational velocity $v_{\rm ZAMS}$ given in Table~\ref{omc_ome}.}
\label{omz_omt}       
\end{figure}

  Following the question/hypothesis (\ref{Q2}), we ascribed several values to $p$ and asked which $J/M$ values explain $\ensuremath{v}=231$\,\ensuremath{\mbox{km}\,\mbox{s}^{-1}}\ at $\ensuremath{t/t_\mathrm{MS}}=0.3$, if the $2.5\,\ensuremath{M_\odot}$ test star rotates with the law given by Eq.\,\ref{omega_r}. The obtained values $J/M=f(p)$ and ratios
$(\Omega_\mathrm{o}/\Omega_{\rm e})$ are given in the upper part of Table~\ref{omc_ome} under
the label ``(\ref{Q2})''. We then required that the $p_{\rm ZAMS}$ values according to the $J/M=f(p)$ values obtained can account for the specific rotational velocity $\ensuremath{v}=182$\,\ensuremath{\mbox{km}\,\mbox{s}^{-1}}\ on the ZAMS (lower part of Table\,\ref{omc_ome} under the label ``(\ref{Q2})''). From Table~\ref{omc_ome} it is clear that the higher the value of the contrast parameter $p$, the higher $J/M$ needs to be to explain the same average equatorial velocity. We also notice that as soon as $p_{\rm ZAMS}\gtrsim0.53$, $J/M$ exceeds $(J/M)_{\rm ZAMS}^{\rm crit}=1.97\times10^{17}$\,cm$^2$\,s$^{-1}$, so that the star must be a neat differential rotator already on the ZAMS. In Fig.~\ref{omz_omt} we show the rotation laws sketched with Eq.\,\ref{omega_r} that would correspond to the parameters $p$ and $(\Omega_\mathrm{o}/\Omega_{\rm e})$ given in Table~\ref{omc_ome} for the ZAMS and for $\ensuremath{t/t_\mathrm{MS}}=0.36$.
 
\begin{table}[!htp]
\caption[]{Contrast parameters $p$, angular velocity ratios $\Omega_\mathrm{o}/\Omega_{\rm e}$
and total specific angular momenta $J/M$ in a $2.5\,\ensuremath{M_\odot}$ test star.}
\centering
\tabcolsep 4.0pt
\begin{tabular}{l|cccccc}
\hline\hline	
 & \multicolumn{4}{c}{Questions/hypotheses} & \multicolumn{2}{c}{}\\
$t/t_{\rm MS}=0.36$ & $\overbrace{\color{white}..}^{\mbox{\small (\ref{Q1})}}$ &  \multicolumn{5}{c}{$\overbrace{\color{white}....................................................}^{\mbox{\small (\ref{Q2})}}$} \\
$p$                                           & 0.00 & 0.20 & 0.40 & 0.60 & 0.80 & 1.00 \\
$\Omega_\mathrm{o}/\Omega_{\rm e}$                     & 1.00 & 1.21 & 1.52 & 2.06 & 3.19 & 7.02 \\ 
$(J/M)/10^{17}{\rm cm^2\,s^{-1}}$               & 1.27 & 1.48 & 1.81 & 2.36 & 3.52 & 7.47 \\
\hline 
$t/t_{\rm MS}=0.00$  & $\overbrace{\color{white}..}^{\mbox{\small (\ref{Q1})}}$ &  \multicolumn{5}{c}{$\overbrace{\color{white}....................................................}^{\mbox{\small (\ref{Q2})}}$} \\
$p_{\rm ZAMS}$                                & 0.25 & 0.39 & 0.53 & 0.67 & 0.81 & 0.96 \\
$(\Omega_\mathrm{o}/\Omega_{\rm e})_{\rm ZAMS}$        & 1.31 & 1.56 & 1.96 & 2.63 & 4.03 & 8.80 \\ 
$(J/M)/10^{17}{\rm cm^2\,s^{-1}}$               & 1.27 & 1.48 & 1.81 & 2.36 & 3.52 & 7.47 \\
\hline 
\end{tabular}
\label{omc_ome}
\end{table}

\section{Comments and conclusions}
\label{disc_concl}

Paper~III provides rotational velocity distributions for A-type star groups defined according to their spectral types. The derivation of fundamental parameters allowed us to disentangle the effects of mass and evolutionary stage in these distributions. In spite of this more refined classification, the striking features already observed in Paper~III remain unchanged (Figs.\,\ref{distrib2} and \ref{distrib3}): (i) the noticeable lack of slow rotators in the low-mass end, i.e. $M\lesssim2.5\ensuremath{M_\odot}$ (ii) the genuine bimodal distributions for the high-mass end ($M\gtrsim2.5\ensuremath{M_\odot}$).
 Taking advantage of the results shown in Figs.~\ref{distrib-age}, \ref{vel_crit}, and \ref{om_rig_dif_obs}, we can attempt to put forward the following comments on some properties of these distributions.

   On the one hand, we saw in Figs.~\ref{vel_crit} and \ref{om_rig_dif_obs} that in intermediate and late A-type stars ($1.6\lesssim M/\ensuremath{M_\odot}\lesssim2.5$) the rotational velocities undergo an acceleration from $t/t_{\rm MS}\approx0$ to $t/t_{\rm MS}\approx0.3$--0.4 and then remain high for a fairly long time, which can certainly help to keep the fraction of slow rotators low during subsequent evolutionary stages up to the TAMS. 

  On the other hand, Fig.~\ref{om_rig_dif_obs} shows that stars with masses $2.5\lesssim M/\ensuremath{M_\odot}\lesssim 3.5$ also accelerate their surface rotation in the time interval ranging from $t/t_{\rm MS}\approx0$ to $t/t_{\rm MS}\approx0.3$, but  undergo an efficient deceleration up to the TAMS, which we already commented in Sects.~\ref{obs_vel_evol} and \ref{obs_vel_evoldet}. We could then think of an initial, more or less genuine, bimodal distribution of rotational velocities among stars with masses in the $2.5\lesssim M/\ensuremath{M_\odot}\lesssim 3$ interval. The commented deceleration of rotational velocities can help to renew the population of slowly rotating stars as the evolution proceeds.

   We conclude, however, that we have no substantial evidence to say why stars in the low-mass interval are more or less systematically not subject to efficient enough decelerating mechanisms that enable them to start their MS evolutionary phase with fairly high rotational velocities. In the same way the fraction of ZAMS slow rotators remains unexplained as well. We can guess, nevertheless, at two possibilities: (i) some stars in the present high-mass interval could start their MS phase after having completely exhausted their angular momentum in the pre-MS phase through magnetic braking, since it is difficult to believe that they never had any angular momentum \citep{Tal78}; (ii) there could be a fraction of undetected binaries where the tidal braking could have some effect. A systematic search of the fraction of rapidly rotating stars among those with small \ensuremath{v\sin i}\ parameters could perhaps shed new light on this question.

  We have studied the evolution of the ratio of average true rotational velocities $\langle\ensuremath{v/v_\mathrm{crit}}\rangle$ (\ensuremath{v}\ is the true equatorial velocity; \ensuremath{v_\mathrm{crit}}\ is the critical equatorial velocity) of MS stars with masses ranging from $1.7\,\ensuremath{M_\odot}$ to $3.3\,\ensuremath{M_\odot}$. Using 2D models of rotating stars we calculated the \ensuremath{v_\mathrm{crit}}\ for individual masses and ages, and obtained the evolution of the average true equatorial velocity $\langle\ensuremath{v}\rangle$ for three stellar masses: 2, 2.5 and 3\,\ensuremath{M_\odot}\ plotted  in Figs.~\ref{vel_crit} and 
\ref{om_rig_dif_obs}. These masses summarize the main characteristics of the evolution of rotational velocities in the MS of stars in the $1.7\leq M/\ensuremath{M_\odot}\leq3.3$  mass range.

  The observed evolution of rotational velocities strongly differs from that theoretically predicted for two limiting cases of internal angular momentum redistribution: (i) rigid rotation (total redistribution); (ii) conservation of the specific angular momentum by each stellar shell (no redistribution). For all studied masses, the observed trends suggest that some effective mechanism of angular momentum redistribution favors the acceleration of the rotation in the stellar surface during the first third of the MS phase. This also makes stars in the studied mass-range possibly behave as differential rotators during all their MS phase. After the acceleration in the first third of the MS phase, stars with masses $M=2\,\ensuremath{M_\odot}$ evolve with little change of their equatorial rotational velocity in the second third of the MS. In the last third of the MS, the rotational velocities decelerate because they were conserving the specific angular momentum per shell without any redistribution. Stars with masses $M\ge2.5\,\ensuremath{M_\odot}$ seem to have an efficient deceleration of the surface rotation during the second third of the MS, which is faster than imposed by a regime of no redistribution of angular momentum. A new acceleration of the surface rotational velocity seems to happen in the $0.6\lesssim t/t_{\rm MS}\lesssim0.7$ time interval, but it decelerates again in the last third of the MS phase. This might be caused by some redistribution of angular momentum toward the center of stars, since according to their low mass-loss rates they cannot undergo strong losses of angular momentum.
  The variation of the surface rotational velocities proceeds at characteristic time scales that are mass-dependent, i.e. $\delta t\approx 0.2\,t_{\rm MS}$.

  The comparison of the observed evolution of rotational velocities for stars with $M=3\,\ensuremath{M_\odot}$ with that predicted by detailed calculations of redistribution processes in the stellar interiors shown in Fig.~\ref{obs_emmb} reveals very strong deviations between theory and observations in the first half of the MS evolutionary phase. 
 We note that a more pronounced spin-down trend during the second part of the MS compared with the $M=3\,\ensuremath{M_\odot}$ model from \citet{Ekm_08} is found by HGM. According to the present work, this faster spin-down seems to be present in the last third of the MS, as shown in Fig\,\ref{obs_emmb}b.

  From the total specific angular momentum calculated to account for the observed velocities of stars in different evolutionary stages if they were rigid rotators, we conclude that stars of all masses studied here could start the MS evolutionary phase with a lower total angular momentum  than the limiting value ascribed to critical rigid rotators on the ZAMS. Also, the observed behavior of the equatorial velocities suggests that they evolve during most of the MS phase as simple differential rotators. However, the angular momenta obtained for masses $M\ge 2.5\,\ensuremath{M_\odot}$ in their last third of the MS phase suggest that they must have larger amounts of rotational energy than rigid critical rotators can bear in the TAMS. This indicates that by the end of the MS they evolve as neat differential rotators.

  Using a test star with $M=2.5\,\ensuremath{M_\odot}$, we have speculated on the properties that its internal rotation may have on the ZAMS, if at this stage the surface rotational velocity corresponded to the total angular momentum required at $t/t_{\rm MS}=0.36$, which is the highest estimated for stars of this mass and in particular, it is higher than the expected one on the ZAMS behaving as rigid rotator. We conclude that an angular velocity ratio $\Omega_\mathrm{o}/\Omega_{\rm e}$ up to 1.3 on the ZAMS could be expected. Assuming that the $2.5\,\ensuremath{M_\odot}$ test star is a simple differential rotator ($J/M<(J/M)^\mathrm{rigid}_\mathrm{crit}$), its equatorial velocity can correspond to angular velocity ratios $\Omega_\mathrm{o}/\Omega_{\rm e}$ as high as 2.6 on the ZAMS, so that $\Omega_\mathrm{o}/\Omega_{\rm e}\lesssim2.1$ for $p=0.6$ at $t/t_{\rm MS}=0.36$. If the object were a neat differential rotator ($J/M>(J/M)^\mathrm{rigid}_\mathrm{crit}$), stable $\Omega_\mathrm{o}/\Omega_{\rm e}$ ratios up to 8.8 could exist on the ZAMS and up to 7 at $t/t_{\rm MS}=0.36$.

   The present discussion assumes that the internal rotation of stars has a shellular-like distribution. We noted in Paper~III, however, that some A-type stars could present a radiative/convective structural dichotomy that could induce other types of angular velocity laws in the external layers and could accordingly affect the interpretation of the measured rotational velocities. This dichotomy might be responsible for anomalous von Zeipel coefficients in fast rotating A-type stars \citep{Zho_09}. Fast rotation can, in particular, affect the temperature gradient in the stellar envelopes, producing thus enlarged convective zones in a similar way as was suggested for massive fast rotators \citep{Mar_08}. Convection and rotation may then be coupled somehow to induce other rotational laws than of mere shellular type \citep{Zoc_11a,Zoc_11b}. In the future, tests should then be made to see which internal rotation profiles can actually prevail. These tests could perhaps be carried out by studying the non-radial pulsation modes of early A-type stars.

\begin{acknowledgements} 
We are thankful to an anonymous referee for her/his valuable comments and suggestions that helped to improve the presentation of our results. We are also very grateful to Prof. Hugo Levato for the communication of his data. We thank Mrs A.~Peter for her prompt and efficient language editing of this paper.
This work has widely made use of the CDS database.

\end{acknowledgements}


\Online
\begin{appendix}
\section{Models of rigid rotators}
\label{mod_rig}

  Only the general dynamical aspects induced by rotation were considered here to calculate the mass distribution in a star and consequently the gravitational potential of a centrifugally distorted star. We separated the primary dynamical effects produced by rotation from those induced by evolution. The primary thermodynamic effects carried by the stellar evolution were taken into account using barotropic relations calculated with stellar models without rotation. We assumed therefore that the changes produced in the $P=P(\rho)$ relation at a given evolutionary stage of a star by the several instabilities and the diffusion of chemical elements unleashed through the stellar evolution by rotation have second-order effects on the establishment of the dynamical equilibrium of the rotating star. In principle, one could use the barotropic relations derived with models of stellar evolution with shellular rotation, but the results will not be more reliable. This approach is used and discussed recently by \citet{Zoc_11b} for massive and intermediate-mass fast rotating stars.

 Because we are not interested in the precise description of all non-linear time-dependent phenomena associated with the viscosity and with internal flows in rotating stars, and because the total energy carried by the meridional circulation is low, our models are axisymmetric, steady state and circulation free. Because of these assumptions, our model-stars behave as barotropes [Poincar\'e-Wavre theorem \citep{Tal78}]. We adopted internal rotational laws of conservative form, $\Omega\!=\!\Omega(\varpi)$, where $\varpi$ is the distance to the rotation axis. The rigid rotation is a special case of this type of rotational laws. In this case, the gravitational potential $\Phi(\varpi,z)$ and the density  distribution $\rho(\varpi,z)$ in the rotating star are simultaneous solutions to the hydrostatic equilibrium equation:
\begin{equation}
\rho^{-1}\nabla P = \nabla\Phi+j^2\varpi^{-3}\hat{e}_{\varpi},
\label{hydeq}
\end{equation}
and to the Poisson equation
\begin{equation}
\Delta\Phi = 4\pi G\rho,
\label{poiss}
\end{equation}
 where ($\varpi,\phi,z$) are the cylindrical coordinates with $z$ containing the rotation axis; $\hat{e}_{\varpi}$ is the unit vector perpendicular to the $z$-axis; $P$ is the pressure; $j=\Omega\varpi^2$ is the specific angular momentum.

  Equations (\ref{hydeq}) and (\ref{poiss}) are solved with the adopted complementary barotropic relation
\begin{equation}
P(\rho) = a\rho^{\gamma_a}+b\rho^{\gamma_b}\ ,
\label{twobarot}
\end{equation}
 where the constants $a$, $b$, $\gamma_a$ and $\gamma_b$ were adjusted to (i) reproduce the pressure $P_{\rm c}$ and the density $\rho_{\rm c}$ in the center of the non-rotating star of given mass and evolutionary stage; (ii) ensure a continuous distribution of the pressure-density relation at the radius of the stellar core; (iii) obtain the right stellar mass at the stellar radius as tabulated by \citet{Scr_92} for 1D evolutionary models for the initial metallicity $Z=0.02$. The function (\ref{twobarot}) is continued in the stellar atmosphere by another pressure-density relation calculated by \citet{CaiKuz03} for stellar atmospheres as a function of the parameters ($T_{\rm eff},\log g$).

 The first-order effects from the stellar evolution are thus accounted for by the pressure-density relations in the center of the star and by the $\partial P/\partial\rho$ gradients. An additional term in relation (\ref{twobarot}) could in principle also take into account the presence of the convective regions in the envelope induced by fast rotation, but we did not do this here. The only rotational effect considered here on the $P\!=\!P(\rho)$ relation is through the mass-compensation effect \citep{Sam70}, which increases the density $\rho_{\rm c}$ in the center of the star. For this, we iterated $\rho_{\rm c}$ until the nominal stellar mass $M$ was obtained. This iteration also implies that the central pressure $P_{\rm c}$ changes in accordance.

  The gravitational potential $\Phi(\varpi,z)$ was obtained by solving Poisson equation (\ref{poiss}) with the cell-method adapted by \citet{Clt74} for stellar structure calculations. The density distribution $\rho(\varpi,z)$ was derived through the integrated form of (\ref{hydeq}).

  Given an angular rotational velocity $\Omega$ and a barotropic relation (\ref{twobarot}), the solutions of equations (\ref{hydeq}) and (\ref{poiss}) were performed over the entire space. The iteration of $\Phi$ and $\rho$ was stopped when the highest density difference in the ($\varpi,z$)-space is $\mathrm{max}(\delta\rho/\rho)\!\la\!10^{-6}$. In our iterations the virial relation $\delta=[2(K+U)-W]/|W|=0$ ($K$ = kinetic energy; $U$ = internal energy; $W$ = total gravitational potential energy) is verified to better than $\delta\approx2\times10^{-4}$ in the ZAMS models and $\delta\approx6\times10^{-3}$ by the TAMS models. Since in the frame of conservative rotational laws the surfaces of constant pressure, density, and of total potential are parallel, the rotationally distorted shape of our models is defined by the total equipotential surface that contains the polar ``photospheric'' radius $R_{\rm p}$. This radius is identified by the layer whose density satisfies the model-atmosphere relation $\tau_{\rm Ross}(\rho)\!=\!2/3$  in the stellar atmosphere models of \citet{CaiKuz03}. The local effective temperature at the pole needs to also satisfy the gravity darkening effect. We accordingly modified the effective temperature given by \citet{Scr_92} for the given mass $M$ using von Zeipel's approximation \citep{vZl24}. The transformation to the rotation dependent effective temperature was performed following the procedure given in \citet{Frt_05a}.

  The  models calculated in this work are for specific total angular momentum scaled as $J/M=0.1$, 1/3, 2/3 and $0\times(J/M)_{\rm crit}^{\rm TAMS}$, where $(J/M)_{\rm crit}^{\rm TAMS}$ is the critical value supported by the stars on the TAMS. We also assumed that the stars evolve conserving their initial angular momentum. The velocity ratios for $J/M=$ 0 and $1\times(J/M)_{\rm crit}^{\rm TAMS}$ are plotted in Fig.~\ref{om_rig_diff}. The characteristic parameters of the models calculated are given in Table~\ref{param_mod}.

\section{Models of differential rotators}
\label{mod_diff}

\begin{figure}[!htp]
\centering
\resizebox{\hsize}{!}{\includegraphics{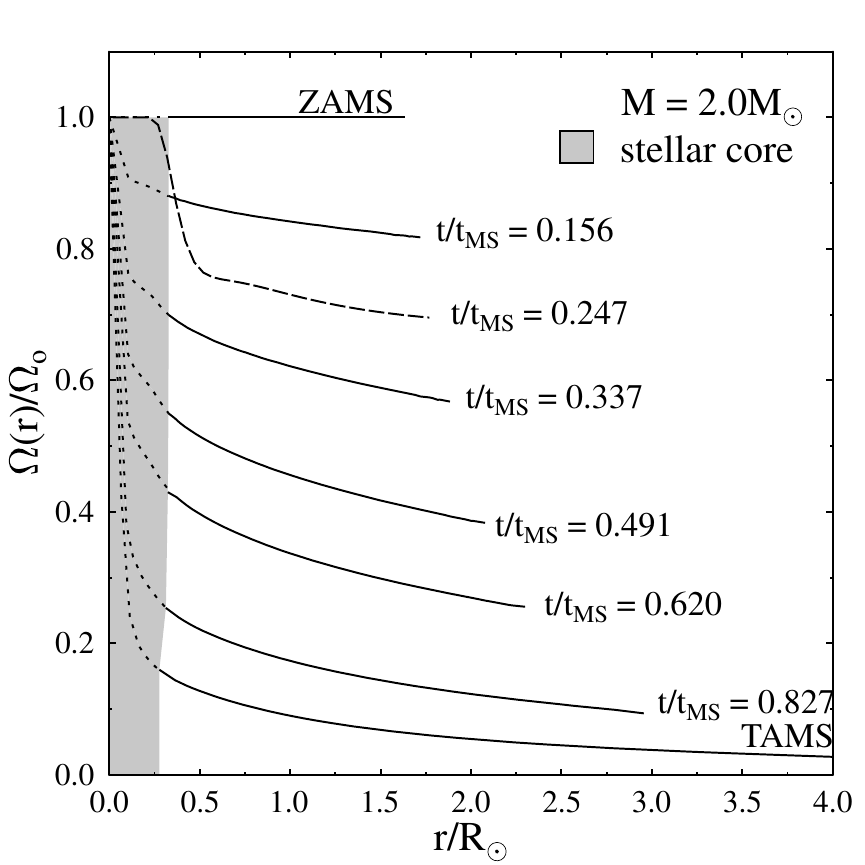}}
\caption{Internal angular velocity at different evolutionary stages in MS for a $2\,\ensuremath{M_\odot}$ rigidly rotating star on the ZAMS that evolves without any exchange of the angular momentum among the stellar shells. The shaded zone represents the stellar core, where the current models with rotation assume that it rotates rigidly. In this case, the internal angular velocity is redistributed as sketched for the $t/t_{\rm MS}\!=\!0.247$ phase.}
\label{omega_norm}       
\end{figure}

   The angular velocity distribution $\Omega'(r')$ at time $t'$ can be derived at time $t$ if we obtain the relation $r' = r'(r)$ that describes the change of the radius $r$ of a shell at time $t$ to the radius $r'$ of the same shell at time $t'$. This can be done observing that the mass inside a given radius $r$ at time $t$ must be the same as that inside $r'$ at time $t'$:
\begin{equation}
\int_0^{r'}\rho(x')x'^2\mbox{d}x' =  \int_0^{r}\rho(x)x^2\mbox{d}x.
\label{mass_consv}
\end{equation}
A similar relation can be obtained from (\ref{veq_difj}) by imposing that
the angular momentum inside $r$ at time $t$ is the same as inside $r'$ at the time $t'$. This relation is much more difficult to manage, however, since the distribution of the angular velocity $\Omega(x')$ and $t'$ must be iterated. In Fig.~\ref{omega_norm} we show the internal rotational profiles at different evolutionary phases identified by the ratios $t/t_{\rm MS}$, where $t$ is the stellar age from the ZAMS and $t_{\rm MS}$ is the stellar age on the TAMS. These distributions were obtained assuming that on the ZAMS the star is a rigid rotator. In Fig.~\ref{omega_norm} the angular velocities are normalized to the angular velocity at $r=0$. The shaded region corresponds to the extent of the stellar core. We note that in current models of massive and intermediate-mass rotating stars, it is assumed that in the convective core $\Omega_{\rm core}\!=~constant$. In that case a distribution like the one sketched for $t/t_{\rm MS}=0.247$ is expected.

 To include the effect of the internal rotation on the stellar moment of inertia and on the stellar radius, we used the angular velocities shown in Fig.~\ref{omega_norm} that are interpreted as conservative, i.e. written as a function of the distance $\varpi$ to the rotational axis. Since the total rotational energy is low, i.e.  $J/M\!\leq\!(J/M)_{\rm crit}$, differences in the estimation of stellar radii are small as noted by the loci of points in Fig.~\ref{om_rig_diff} for $J/M\!=\!0$ and $J/M\!=\!(J/M)_{\rm crit}$.

\setlongtables
\onecolumn
\begin{longtable}{ccccccccc}
\caption{\label{param_mod}Parameters characterizing the models of rigid rotators that evolve with constant angular momentum $J\!=\!q\times{J}_{\rm crit}^{\rm TAMS}$, with $J_{\rm crit}^{\rm TAMS}$ for rigid rotation.}\\
\hline\hline	
\centering
$t/t_{\rm MS}$ & $\Omega$ &  $\Omega/\Omega_{\rm crit}$ & $\rho_{\rm c}$ & 
$R_{\rm p}/R_{\odot}$ & $R_{\rm e}/R_{\rm p}$ & \ensuremath{v} & \ensuremath{v_\mathrm{crit}}\ & $K/|W|$ \\
               & (s$^{-1}$) &                         &  (g\,cm$^{-3}$)    &
                      &                       & (\ensuremath{\mbox{km}\,\mbox{s}^{-1}}) &  \ensuremath{v/v_\mathrm{crit}}\    & \\
\hline
\endfirsthead
\caption{continued.}\\
\hline\hline	
$t/t_{\rm MS}$ & $\Omega$ &  $\Omega/\Omega_{\rm crit}$ & $\rho_{\rm c}$ & 
$R_{\rm p}/R_{\odot}$ & $R_{\rm e}/R_{\rm p}$ & \ensuremath{v} & \ensuremath{v_\mathrm{crit}}\ & $K/|W|$ \\
               & (s$^{-1}$) &                         &  (g\,cm$^{-3}$)    &
                      &                       & (\ensuremath{\mbox{km}\,\mbox{s}^{-1}}) &   \ensuremath{v/v_\mathrm{crit}}\   &   \\
\hline	
\endhead
\hline
\endfoot
\multicolumn{2}{l}{$M/M_{\odot}=$ 1.5} & \multicolumn{4}{c}{$J/M=$ 
6.98$\times10^{15}$ cm$^2$\,s$^{-1}$} & \multicolumn{3}{c}{$q= 0.1$} \\
\hline	
  0.016 & 0.151E-04 & 0.057 &  76.04 & 1.420 & 1.001 &  14.9 & 380.2 & 0.14E-04 \\
  0.335 & 0.142E-04 & 0.063 &  85.51 & 1.577 & 1.001 &  15.6 & 360.8 & 0.13E-04 \\
  0.619 & 0.126E-04 & 0.070 &  92.69 & 1.823 & 1.001 &  16.0 & 335.6 & 0.12E-04 \\
  0.827 & 0.114E-04 & 0.078 & 104.00 & 2.103 & 1.001 &  16.6 & 312.4 & 0.11E-04 \\
  0.920 & 0.111E-04 & 0.085 & 115.62 & 2.259 & 1.001 &  17.5 & 301.4 & 0.11E-04 \\
  1.000 & 0.121E-04 & 0.103 & 147.66 & 2.408 & 1.001 &  17.6 & 291.8 & 0.11E-04 \\
\hline	
\multicolumn{2}{l}{$M/M_{\odot}=$ 1.5} & \multicolumn{4}{c}{$J/M=$ 
3.49$\times10^{16}$ cm$^2$\,s$^{-1}$} & \multicolumn{3}{c}{$q= 0.5$} \\
\hline	
   0.016 & 0.752E-04 & 0.283 &  76.20 & 1.416 & 1.015 &  75.2 & 0.198 & 0.34E-03 \\
  0.335 & 0.709E-04 & 0.314 &  85.69 & 1.572 & 1.018 &  79.0 & 0.219 & 0.32E-03 \\
  0.619 & 0.628E-04 & 0.348 &  92.89 & 1.817 & 1.024 &  81.3 & 0.242 & 0.30E-03 \\
  0.827 & 0.566E-04 & 0.390 & 104.21 & 2.096 & 1.028 &  84.9 & 0.272 & 0.28E-03 \\
  0.920 & 0.553E-04 & 0.424 & 115.86 & 2.251 & 1.034 &  89.5 & 0.297 & 0.27E-03 \\
  1.000 & 0.592E-04 & 0.502 & 148.93 & 2.398 & 1.048 & 103.7 & 0.355 & 0.27E-03 \\
\hline	
\multicolumn{2}{l}{$M/M_{\odot}=$ 1.5} & \multicolumn{4}{c}{$J/M=$ 
6.98$\times10^{16}$ cm$^2$\,s$^{-1}$} & \multicolumn{3}{c}{$q= 1.0$} \\
\hline	
  0.016 & 0.150E-03 & 0.562 &  76.69 & 1.403 & 1.065 & 155.5 & 0.409 & 0.13E-02 \\
  0.335 & 0.141E-03 & 0.623 &  86.24 & 1.558 & 1.081 & 164.8 & 0.457 & 0.13E-02 \\
  0.619 & 0.124E-03 & 0.689 &  93.46 & 1.800 & 1.104 & 172.2 & 0.513 & 0.12E-02 \\
  0.827 & 0.112E-03 & 0.770 & 104.86 & 2.076 & 1.142 & 184.7 & 0.591 & 0.11E-02 \\
  0.920 & 0.109E-03 & 0.837 & 116.58 & 2.229 & 1.189 & 201.2 & 0.667 & 0.11E-02 \\
  1.000 & 0.116E-03 & 1.000 & 149.93 & 2.372 & 1.519 & 291.8 & 1.000 & 0.11E-02 \\
\hline	
\multicolumn{2}{l}{$M/M_{\odot}=$ 1.5} & \multicolumn{4}{c}{$J/M=$ 
1.05$\times10^{17}$ cm$^2$\,s$^{-1}$} & \multicolumn{3}{c}{$q= 1.5$} \\
\hline	
  0.016 & 0.222E-03 & 0.836 &  77.38 & 1.387 & 1.152 & 247.3 & 0.651 & 0.30E-02 \\
  0.157 & 0.220E-03 & 0.885 &  82.41 & 1.453 & 1.140 & 253.1 & 0.681 & 0.29E-02 \\
  0.335 & 0.209E-03 & 0.927 &  86.87 & 1.544 & 1.170 & 263.2 & 0.729 & 0.28E-02 \\
  0.489 & 0.197E-03 & 0.979 &  90.72 & 1.665 & 1.193 & 272.1 & 0.783 & 0.27E-02 \\
\hline	
\multicolumn{2}{l}{$M/M_{\odot}=$ 2.0} & \multicolumn{4}{c}{$J/M=$ 
7.25$\times10^{15}$ cm$^2$\,s$^{-1}$} & \multicolumn{3}{c}{$q= 0.1$} \\
\hline	
  0.010 & 0.116E-04 & 0.047 &  60.26 & 1.625 & 1.000 &  13.1 & 403.8 & 0.95E-05 \\
  0.337 & 0.101E-04 & 0.051 &  60.11 & 1.872 & 1.001 &  13.2 & 376.2 & 0.87E-05 \\
  0.620 & 0.847E-05 & 0.058 &  62.24 & 2.286 & 1.001 &  13.5 & 340.5 & 0.77E-05 \\
  0.827 & 0.725E-05 & 0.070 &  68.71 & 2.863 & 1.001 &  14.5 & 304.2 & 0.69E-05 \\
  0.946 & 0.638E-05 & 0.071 &  75.96 & 3.564 & 1.001 &  15.8 & 290.5 & 0.63E-05 \\
  1.000 & 0.710E-05 & 0.103 &  98.85 & 3.744 & 1.001 &  16.1 & 266.2 & 0.66E-05 \\
\hline	
\multicolumn{2}{l}{$M/M_{\odot}=$ 2.0} & \multicolumn{4}{c}{$J/M=$ 
3.62$\times10^{16}$ cm$^2$\,s$^{-1}$} & \multicolumn{3}{c}{$q= 0.5$} \\
\hline	
  0.010 & 0.579E-04 & 0.234 &  60.35 & 1.621 & 1.010 &  65.9 & 0.163 & 0.24E-03 \\
  0.337 & 0.504E-04 & 0.253 &  60.20 & 1.868 & 1.012 &  66.3 & 0.176 & 0.22E-03 \\
  0.620 & 0.423E-04 & 0.289 &  62.32 & 2.281 & 1.015 &  68.1 & 0.200 & 0.19E-03 \\
  0.827 & 0.362E-04 & 0.348 &  68.80 & 2.856 & 1.022 &  73.4 & 0.241 & 0.17E-03 \\
  0.946 & 0.318E-04 & 0.352 &  76.06 & 3.556 & 1.032 &  81.2 & 0.280 & 0.16E-03 \\
  1.000 & 0.348E-04 & 0.502 &  99.29 & 3.732 & 1.047 &  94.5 & 0.356 & 0.16E-03 \\
\hline	
\multicolumn{2}{l}{$M/M_{\odot}=$ 2.0} & \multicolumn{4}{c}{$J/M=$ 
7.25$\times10^{16}$ cm$^2$\,s$^{-1}$} & \multicolumn{3}{c}{$q= 1.0$} \\
\hline	
  0.010 & 0.115E-03 & 0.465 &  60.61 & 1.613 & 1.041 & 134.8 & 0.334 & 0.94E-03 \\
  0.337 & 0.100E-03 & 0.505 &  60.45 & 1.857 & 1.049 & 136.0 & 0.362 & 0.86E-03 \\
  0.620 & 0.840E-04 & 0.574 &  62.57 & 2.267 & 1.064 & 141.1 & 0.414 & 0.77E-03 \\
  0.827 & 0.717E-04 & 0.690 &  69.08 & 2.837 & 1.102 & 156.1 & 0.513 & 0.69E-03 \\
  0.946 & 0.629E-04 & 0.696 &  76.37 & 3.530 & 1.181 & 182.5 & 0.628 & 0.62E-03 \\
  1.000 & 0.684E-04 & 1.000 &  99.65 & 3.703 & 1.512 & 266.2 & 1.000 & 0.64E-03 \\
\hline	
\multicolumn{2}{l}{$M/M_{\odot}=$ 2.0} & \multicolumn{4}{c}{$J/M=$ 
1.09$\times10^{17}$ cm$^2$\,s$^{-1}$} & \multicolumn{3}{c}{$q= 1.5$} \\
\hline	
  0.010 & 0.172E-03 & 0.695 &  60.91 & 1.598 & 1.077 & 206.4 & 0.511 & 0.21E-02 \\
  0.156 & 0.164E-03 & 0.716 &  60.93 & 1.680 & 1.087 & 208.3 & 0.529 & 0.20E-02 \\
  0.337 & 0.150E-03 & 0.753 &  60.74 & 1.842 & 1.094 & 210.0 & 0.558 & 0.19E-02 \\
  0.490 & 0.137E-03 & 0.800 &  61.41 & 2.033 & 1.101 & 213.3 & 0.596 & 0.18E-02 \\
  0.620 & 0.125E-03 & 0.854 &  62.88 & 2.249 & 1.135 & 222.2 & 0.653 & 0.17E-02 \\
  0.732 & 0.116E-03 & 0.928 &  65.63 & 2.501 & 1.149 & 231.4 & 0.716 & 0.16E-02 \\
\hline	
\multicolumn{2}{l}{$M/M_{\odot}=$ 2.5} & \multicolumn{4}{c}{$J/M=$ 
8.93$\times10^{15}$ cm$^2$\,s$^{-1}$} & \multicolumn{3}{c}{$q= 0.1$} \\
\hline	
  0.012 & 0.107E-04 & 0.046 &  46.66 & 1.821 & 1.000 &  13.6 & 423.4 & 0.10E-04 \\
  0.342 & 0.907E-05 & 0.050 &  44.77 & 2.135 & 1.001 &  13.5 & 391.0 & 0.90E-05 \\
  0.629 & 0.756E-05 & 0.057 &  45.82 & 2.617 & 1.001 &  13.8 & 353.2 & 0.80E-05 \\
  0.835 & 0.642E-05 & 0.069 &  50.00 & 3.306 & 1.001 &  14.8 & 314.2 & 0.72E-05 \\
  0.949 & 0.591E-05 & 0.085 &  58.34 & 4.017 & 1.001 &  16.5 & 285.1 & 0.67E-05 \\
  1.000 & 0.619E-05 & 0.103 &  71.66 & 4.414 & 1.001 &  16.5 & 273.3 & 0.68E-05 \\
\hline	
\multicolumn{2}{l}{$M/M_{\odot}=$ 2.5} & \multicolumn{4}{c}{$J/M=$ 
4.46$\times10^{16}$ cm$^2$\,s$^{-1}$} & \multicolumn{3}{c}{$q= 0.5$} \\
\hline	
  0.012 & 0.536E-04 & 0.231 &  46.74 & 1.816 & 1.009 &  68.4 & 0.162 & 0.25E-03 \\
  0.342 & 0.453E-04 & 0.250 &  44.84 & 2.129 & 1.011 &  67.9 & 0.174 & 0.23E-03 \\
  0.629 & 0.377E-04 & 0.284 &  45.88 & 2.611 & 1.015 &  69.6 & 0.197 & 0.20E-03 \\
  0.835 & 0.320E-04 & 0.344 &  50.07 & 3.298 & 1.021 &  75.1 & 0.239 & 0.18E-03 \\
  0.949 & 0.295E-04 & 0.425 &  58.43 & 4.007 & 1.032 &  84.8 & 0.298 & 0.17E-03 \\
  1.000 & 0.304E-04 & 0.506 &  72.05 & 4.402 & 1.047 &  97.4 & 0.358 & 0.17E-03 \\
\hline	
\multicolumn{2}{l}{$M/M_{\odot}=$ 2.5} & \multicolumn{4}{c}{$J/M=$ 
8.93$\times10^{16}$ cm$^2$\,s$^{-1}$} & \multicolumn{3}{c}{$q= 1.0$} \\
\hline	
  0.012 & 0.107E-03 & 0.460 &  46.95 & 1.808 & 1.040 & 139.8 & 0.330 & 0.10E-02 \\
  0.342 & 0.902E-04 & 0.497 &  45.03 & 2.119 & 1.046 & 139.2 & 0.356 & 0.90E-03 \\
  0.629 & 0.751E-04 & 0.565 &  46.07 & 2.595 & 1.061 & 143.9 & 0.407 & 0.80E-03 \\
  0.835 & 0.635E-04 & 0.683 &  50.27 & 3.277 & 1.096 & 158.8 & 0.505 & 0.71E-03 \\
  0.949 & 0.583E-04 & 0.841 &  58.66 & 3.978 & 1.178 & 189.9 & 0.666 & 0.66E-03 \\
  1.000 & 0.598E-04 & 1.000 &  72.37 & 4.365 & 1.512 & 273.3 & 1.000 & 0.65E-03 \\
\hline	
\multicolumn{2}{l}{$M/M_{\odot}=$ 2.5} & \multicolumn{4}{c}{$J/M=$ 
1.34$\times10^{17}$ cm$^2$\,s$^{-1}$} & \multicolumn{3}{c}{$q= 1.5$} \\
\hline	
  0.012 & 0.160E-03 & 0.687 &  47.21 & 1.785 & 1.079 & 214.1 & 0.506 & 0.22E-02 \\
  0.163 & 0.148E-03 & 0.707 &  45.79 & 1.910 & 1.080 & 213.0 & 0.520 & 0.21E-02 \\
  0.342 & 0.135E-03 & 0.742 &  45.26 & 2.098 & 1.093 & 214.8 & 0.549 & 0.20E-02 \\
  0.496 & 0.123E-03 & 0.784 &  45.38 & 2.312 & 1.113 & 219.6 & 0.589 & 0.19E-02 \\
  0.629 & 0.112E-03 & 0.841 &  46.35 & 2.574 & 1.151 & 230.2 & 0.652 & 0.18E-02 \\
  0.742 & 0.102E-03 & 0.913 &  47.82 & 2.875 & 1.163 & 238.3 & 0.713 & 0.17E-02 \\
\hline	
\multicolumn{2}{l}{$M/M_{\odot}=$ 3.0} & \multicolumn{4}{c}{$J/M=$ 
1.09$\times10^{16}$ cm$^2$\,s$^{-1}$} & \multicolumn{3}{c}{$q= 0.1$} \\
\hline	
  0.007 & 0.102E-04 & 0.046 &  37.16 & 1.999 & 1.001 &  14.3 & 441.0 & 0.11E-04 \\
  0.350 & 0.854E-05 & 0.051 &  34.92 & 2.382 & 1.001 &  14.2 & 404.0 & 0.99E-05 \\
  0.643 & 0.708E-05 & 0.058 &  35.40 & 2.920 & 1.001 &  14.4 & 364.8 & 0.88E-05 \\
  0.844 & 0.600E-05 & 0.070 &  38.64 & 3.684 & 1.001 &  15.4 & 324.8 & 0.78E-05 \\
  0.951 & 0.549E-05 & 0.086 &  44.57 & 4.481 & 1.001 &  17.2 & 294.5 & 0.73E-05 \\
  1.000 & 0.566E-05 & 0.102 &  54.71 & 4.923 & 1.002 &  19.4 & 281.0 & 0.73E-05 \\
\hline	
\multicolumn{2}{l}{$M/M_{\odot}=$ 3.0} & \multicolumn{4}{c}{$J/M=$ 
5.44$\times10^{16}$ cm$^2$\,s$^{-1}$} & \multicolumn{3}{c}{$q= 0.5$} \\
\hline	
  0.007 & 0.511E-04 & 0.231 &  37.22 & 1.995 & 1.009 &  71.7 & 0.163 & 0.28E-03 \\
  0.350 & 0.426E-04 & 0.253 &  34.97 & 2.376 & 1.010 &  71.2 & 0.176 & 0.25E-03 \\
  0.643 & 0.353E-04 & 0.287 &  35.46 & 2.912 & 1.016 &  72.7 & 0.199 & 0.22E-03 \\
  0.844 & 0.299E-04 & 0.347 &  38.70 & 3.675 & 1.022 &  78.3 & 0.241 & 0.20E-03 \\
  0.951 & 0.274E-04 & 0.427 &  44.63 & 4.469 & 1.032 &  87.9 & 0.298 & 0.18E-03 \\
  1.000 & 0.282E-04 & 0.507 &  54.79 & 4.910 & 1.046 & 100.8 & 0.359 & 0.18E-03 \\
\hline	
\multicolumn{2}{l}{$M/M_{\odot}=$ 3.0} & \multicolumn{4}{c}{$J/M=$ 
1.09$\times10^{17}$ cm$^2$\,s$^{-1}$} & \multicolumn{3}{c}{$q= 1.0$} \\
\hline	
  0.007 & 0.102E-03 & 0.461 &  37.40 & 1.983 & 1.040 & 146.2 & 0.332 & 0.11E-02 \\
  0.350 & 0.849E-04 & 0.504 &  35.13 & 2.363 & 1.047 & 146.2 & 0.362 & 0.98E-03 \\
  0.643 & 0.703E-04 & 0.571 &  35.61 & 2.895 & 1.062 & 150.3 & 0.412 & 0.87E-03 \\
  0.844 & 0.594E-04 & 0.688 &  38.86 & 3.650 & 1.096 & 165.4 & 0.509 & 0.77E-03 \\
  0.951 & 0.541E-04 & 0.844 &  44.82 & 4.435 & 1.177 & 196.5 & 0.667 & 0.72E-03 \\
  1.000 & 0.555E-04 & 1.000 &  55.03 & 4.869 & 1.495 & 281.0 & 1.000 & 0.71E-03 \\
\hline	
\multicolumn{2}{l}{$M/M_{\odot}=$ 3.0} & \multicolumn{4}{c}{$J/M=$ 
1.63$\times10^{17}$ cm$^2$\,s$^{-1}$} & \multicolumn{3}{c}{$q= 1.5$} \\
\hline	
  0.007 & 0.152E-03 & 0.689 &  37.60 & 1.957 & 1.070 & 221.9 & 0.503 & 0.25E-02 \\
  0.166 & 0.140E-03 & 0.713 &  36.06 & 2.112 & 1.080 & 222.1 & 0.523 & 0.23E-02 \\
  0.350 & 0.127E-03 & 0.753 &  35.30 & 2.336 & 1.090 & 224.6 & 0.556 & 0.22E-02 \\
  0.510 & 0.115E-03 & 0.795 &  35.28 & 2.578 & 1.095 & 226.4 & 0.588 & 0.21E-02 \\
  0.643 & 0.105E-03 & 0.852 &  35.76 & 2.870 & 1.112 & 232.8 & 0.638 & 0.19E-02 \\
  0.755 & 0.958E-04 & 0.929 &  37.03 & 3.221 & 1.148 & 246.4 & 0.715 & 0.18E-02 \\	
\hline
\end{longtable}
\tablefoot{For each mass and for $q=0.1$ is given the \ensuremath{v_\mathrm{crit}}, while for other $p$ values is given the ratio \ensuremath{v/v_\mathrm{crit}}. $t/t_{\rm MS}$ is the fractional stellar age, $t_{\rm MS}$ being the time the star of mass $M$ can spend in the MS. $\Omega$ is the angular velocity and $\Omega_{\rm crit}$ is the critical angular velocity. $\rho_{\rm c}$ is the density at the stellar center. $R_{\rm p}/R_{\odot}$ is the polar radius in solar units and $R_{\rm e}/R_{\rm p}$ is the equatorial to polar radii ratio. $\ensuremath{v}$ is the equatorial rotational velocity and $\ensuremath{v_\mathrm{crit}}$ the equatorial critical rotational velocity. $K/|W|$ is the ratio of kinetic to gravitational potential energy.}

\end{appendix}

\end{document}